\documentclass[usenatbib]{mnras}

\usepackage[english]{babel}
\usepackage{amsmath}
\usepackage{amssymb}
\usepackage{graphicx}
\usepackage{tikz}
\usepackage{lipsum,adjustbox}
\usetikzlibrary{shapes,arrows}
\usepackage{breqn}
\usepackage{appendix}
\usepackage{color}

\newcommand{\as}{\color{black}}

\newcommand{\kms}{km~s\ensuremath{^{-1}}}
\newcommand{\msun}{$M_{\odot}$}
\newcommand{\mh}{$M_{{\rm H}_2}$}

\newcommand{\mstar}{$M_{\ast}$}

\newcommand{\fmol}{$f_{\rm [C\small II\normalsize],mol}$}
\newcommand{\cii}{[C\small II\normalsize]}

\usepackage{etoolbox}
\makeatletter
\newcount\c@additionalboxlevel
\newcount\c@maxboxlevel
\patchcmd\@combinedblfloats{\box\@outputbox}{%
  \stepcounter{additionalboxlevel}%
  \box\@outputbox
}{}{\errmessage{\noexpand\@combinedblfloats could not be patched}}

\AtBeginShipout{%
  \ifnum\value{additionalboxlevel}>\value{maxboxlevel}%
    \typeout{Warning: maxboxlevel might be too small, increase to %
      \the\value{additionalboxlevel}%
    }%
  \fi 
  \@whilenum\value{additionalboxlevel}<\value{maxboxlevel}\do{%
    \typeout{* Additional boxing of page `\thepage'}%
    \setbox\AtBeginShipoutBox=\hbox{\copy\AtBeginShipoutBox}%
    \stepcounter{additionalboxlevel}%
  }%
  \setcounter{additionalboxlevel}{0}%
}
\makeatother

\begin{document}

\title[Disentangling C\small II\Large\space emission on galaxy-wide scales using a unified ISM model.]{Radiative transfer meets Bayesian statistics: where does a galaxy's [C\small II\huge] emission come from?}
\author[G. Accurso et al. ]{G. Accurso$^{1}$\thanks{E-mail:
gioacchino.accurso.13@ucl.ac.uk}, A. Saintonge$^{1}$, T.G. Bisbas$^{2,3}$, S. Viti$^{1}$. \\ 
$^{1}$Dept of Physics and Astronomy, University College London, Gower Street, London, WC1E 6BT, UK \\
$^{2}$Max-Planck-Institut f\"ur Extraterrestrische Physik, Giessenbachstrasse 1, D-85748 Garching, Germany. \\
$^{3}$Department of Astronomy, University of Florida, Gainesville, FL 32611, USA.}

\maketitle

\date{Accepted Date Received Date; in original form Date}

\begin{abstract}
The [C\small II\normalsize] 158$\mu$m emission line can arise in all phases of the ISM, therefore being able to disentangle the different contributions is an important yet unresolved problem when undertaking galaxy-wide, integrated [C\small II\normalsize] observations. We present a new multi-phase 3D radiative transfer interface that couples {\sc starburst99}, a stellar spectrophotometric code, with the photoionisation and astrochemistry codes {\sc mocassin} and {\sc 3d-pdr}. We model entire star forming regions, including the ionised, atomic and molecular phases of the ISM, and apply a Bayesian inference methodology to parametrise how the fraction of the [C\small II\normalsize] emission originating from molecular regions, \fmol, varies as a function of typical integrated properties of galaxies in the local Universe. The main parameters responsible for the variations of \fmol\ are specific star formation rate (SSFR), gas phase metallicity, H\footnotesize II \normalsize region electron number density ($n_e$), and dust mass fraction. For example, \fmol\ can increase from 60\% to 80\% when either $n_e$ increases from 10$^{1.5}$ to 10$^{2.5}$cm$^{-3}$, or SSFR decreases from $10^{-9.6}$ to $10^{-10.6}$ yr$^{-1}$. Our model predicts for the Milky Way that \fmol$=75.8\pm5.9$\%, in agreement with the measured value of 75$\%$. When applying the new prescription to a complete sample of galaxies from the Herschel Reference Survey (HRS), we find that anywhere from 60 to 80\% of the total integrated [C\small II\normalsize] emission arises from molecular regions. 
\end{abstract}


%methods: numerical - radiative transfer - ISM: H\footnotesize II \large regions - ISM: PDR regions - astrochemistry - galaxies.}

%%%%%%%%%%%%%%%%%%%%%%%%%%%%
%%%%%%%%%%%%%%%%%%%%%%%%%%%%
\section{Introduction}
{\as The [C\small II\normalsize] 158$\mu$m emission of singly ionised carbon is one of the strongest cooling lines of the ISM, and can carry up to a few percent of the total far-infrared (FIR) energy emitted from galaxies. It correlates with the total molecular gas mass of galaxies (\mh), as measured from $^{12}$CO emission, and with the total star formation rate inferred from FIR luminosity. For these reasons, [C\small II\normalsize] is an important and widely used tracer of massive star formation in galaxies at both low and high redshifts \citep{2010ApJ...724..957S, 2011MNRAS.416.2712D}. 

Ionised carbon (C$^{+}$) can be found throughout the ISM, from photodissociation regions (PDRs) to diffuse ionised and atomic regions, owing to the fact that carbon has a first ionisation potential of 11.3eV, lower than hydrogen's \citep{1999ApJ...527..795K, 2003MNRAS.346.1055K}. While [C\small II\normalsize] originates in good part from PDRs, explaining the correlation with \mh\ and SFR, observations have shown that a non-negligible fraction of the emission can originate from the ionised and diffuse atomic gas components where massive star formation does not occur \citep{1994ApJ...436..720H,2010HiA....15..408V,2010A&A...521L..17L,2013A&A...554A.103P, 2013A&A...553A.114K}. 

The CO molecule, as a tracer of the cold molecular phase of the ISM, suffers from the opposite problem: in low metallicity environments, the CO molecule can be photo-dissociated by UV radiation while H$_2$ self-shields and survives, resulting in the presence of molecular gas that is missed by CO observations. For example, \citet{2010A&A...521L..18V} find that $\sim$ 25$\%$ of the molecular gas in the Milky Way may be CO-dark in such a way. If it were possible to discern the contribution of the different ISM phases to the total \cii\ emission, then the combination of CO and \cii\ measurements could increase significantly the accuracy of \mh\ calculations \citep{2016A&A...586A..37M}. There is also a new interest in using \cii\ as a probe of the ISM in $z>5$ galaxies and up to well into the epoch of reionisation; such studies are now made increasingly possible with facilities such as ALMA and NOEMA \citep[e.g.][]{ota14,riechers14,maiolino15}. In this context also, disentangling the contributions from the different phases of the ISM is of significant importance, yet this problem remains unsolved.

Solving this problem requires one of two things: either high spatial resolution observations of several FIR lines such as [N\small II\normalsize]122,205$\mu$m and [O\small I\normalsize]63,145$\mu$m in addition to \cii, or a self-consistent model of the ISM on galaxy-wide scales, including PDRs, ionised and neutral diffuse regions. Since such detailed observations are only available for a handful of very nearby galaxies, we focus here on the modelling approach.} Numerical codes treating PDRs have been around for decades and have now grown into complex models capable of solving the thermal balance equations and chemical reactions occurring within these regions \citep{2013ascl.soft03004V}. Some codes have aimed to include all the small scale physics to describe the chemical and thermal processes at work in the gas and grains, while others focus on treating the gas-grain chemistry while approximating other processes. Various codes treating one dimensional PDRs have been developed in the past and only recently three-dimensional codes have emerged which can treat PDRs of an arbitrary density distribution \citep{2012MNRAS.427.2100B}. {\as Furthermore, calculations typically treat the ionised and PDR regions separately \citep{2006A&A...451..917R}, which is problematic for emission lines such as \cii\ that originate in both these phases of the ISM and can lead to overestimations of line intensities and incorrect interpretations of the physical conditions in the ISM (e.g. hydrogen column density and incident ionisation field).}

Photoionisation codes, used to model the H\footnotesize II \normalsize regions, have likewise been around for several decades now. {\as They typically work by solving the equations of radiative transfer while making assumptions concerning spherical symmetry. The earliest H\footnotesize II \normalsize region models contained the basic physics of ionisation, recombination of hydrogen and helium, thermal balance, and the emission of photons from the nebula  \citep{1968IAUS...34..205F}, with subsequent codes having seen the addition of other important physical processes such as charge exchange and dielectric recombination, and the consideration of a wider range of ions.} More recently, three dimensional codes have been developed to handle varying geometries using a Monte-Carlo approach to solve the 3D equations of radiative transport \citep{2004MNRAS.348.1337W}. 

Although numerical models for the individual components of the ISM are aplenty, codes which can simulate all aforementioned phases of the ISM consistently are not so common. One very successful example is {\sc cloudy} \citep{2013RMxAA..49..137F}, a plasma simulation which models the ionisation, chemical and thermal state of the gas that may be exposed to an external radiation field coming from a nearby heating sources such as star clusters. {\as The code works by predicting the spectrum from this non-equilibrium gas and simulating its level populations as well as its ionisation, molecular and thermal states, over a wide range of densities and temperatures. However, {\sc cloudy} is intrinsically a 1D code, and involves assumptions concerning the thermal balance in PDRs. Both of these limitations should be lifted to accurately simulate the entire ISM of star-forming regions and galaxies. Indeed, it has been shown that results from 1D and 3D simulations vary depending on the specific physical conditions \citep{2012MNRAS.420..141E}. An example of a successful attempt at building such a 3D, multi-phase radiative transfer code is {\sc Torus-3dpdr} \citep{2015MNRAS.454.2828B}, a Hydrodynamics and Monte Carlo radiative transfer code.} {\sc Torus-3dpdr} does not use the complexity of {\sc mocassin} to calculate particular photoionisation calculations and therefore won't be as accurate.

{\as In this paper, we present a new modelling interface which combines self-consistently state of the art astrochemistry and photoionisation codes:  {\sc 3d-pdr}, a three-dimensional code for treating PDRs and molecular regions,  {\sc mocassin}, a full 3D Monte Carlo photoionisation code, and {\sc starburst99}, a stellar population synthesis code. This integrated code is used to simulate entire star forming regions, including the ionised, neutral and molecular phases of the ISM, with the aim of parametrising how the fraction of the total \cii\ emission originating from molecular regions, \fmol\, varies as a function of the physical conditions in the ISM. A Bayesian Inference technique is used to solve this complex multi-parameter problem, allowing us to derive a series of prescriptions to calculate the contributions of the different phases of the ISM to the total integrated \cii\ emission in extragalactic sources.

In Sections \ref{overviewofcodes} and \ref{couplingmethod} we provide further technical details concerning the codes and the modelling strategy, while the choice of input parameters is explained in Section \ref{modelparamspace}.  The results of the modelling and of the Bayesian analysis to produce scaling relations for \fmol\ are presented in Sections  \ref{numericalresults} and \ref{applicationstongalaxies}, respectively. These prescriptions are validated through comparisons with observations in Sections \ref{obscomparison}. Finally, we conclude in Section \ref{answertopaper} with a concise presentation of our new prescriptions to infer \fmol\ from galaxy-integrated quantities.}

Throughout this paper we use a standard flat $\Lambda \mbox{CDM}$ cosmology with $H_{0} = 70$ km s$^{-1}$ Mpc$^{-1}$ and the IMF from \citet{2003PASP..115..763C}.

%%%%%%%%%%%%%%%%%%%%%%%%%%%%%%%%%%%%%%%%%%%%%%%%%%%
%%%%%%%%%%%%%%%%%%%%%%%%%%%%%%%%%%%%%%%%%%%%%%%%%%%
\section{Overview of Numerical Codes} \label{overviewofcodes}
We here briefly describe each of the codes used within this work; {\as for more detailed explanations, see the references provided for each code}.

\subsection{The {\sc starburst99} code}
{\sc starburst99} \citep{1999ApJS..123....3L, 2010ApJS..189..309L, 2005ApJ...621..695V, 2013ARA&A..51..393C}, hereafter {\sc sb99}, is a multi-purpose evolutionary synthesis code which models the spectrophotometric properties of unresolved stellar populations, and makes predictions for various observables, such as spectral energy distributions, supernovae rates, and mass loss rates. The code simulates a population of stars based on an input metallicity and stellar initial mass function (IMF), and evolves them across the Hertzprung-Russell (HR) diagram. The code creates a grid covering the HR diagram and integrates over it with weights assigned according to properties of the stellar population to produce the integrated spectrum for the population. {\sc sb99} can thus provide the relation between the stellar mass (M$_{\ast}$) and luminosity (L$_{\ast}$ of the stellar population at any time interval; see \citet{2013ARA&A..51..393C} for a comprehensive review of the synthesis technique.

%At each time interval and for each mass a spectrum is assigned and the integrated spectrum for the population is obtained by summing over these contributions

\subsection{The {\sc mocassin} code}
The numerical code {\sc mocassin} (MOnte CArlo SimulationS of Ionised Nebulae, \citet{2003MNRAS.340.1136E}) is a three-dimensional Monte Carlo (MC) radiative transfer code {\as which operates on a non-uniform cartesian grid}. It was originally intended as a tool to construct realistic gas models of planetary nebulae, but has since evolved to incorporate dust radiative transfer \citep{2005MNRAS.362.1038E} and is now used to simulate ionised gas emission on galaxy-wide scales \citep{2013MNRAS.431.2493K}. Photoionization calculations are performed using an iterative Monte Carlo photon energy packet propagating routine, based on the methods presented by \citet{1999A&A...344..282L}. Photons are emitted from the ionising source in random but isotropic directions, and propagate for a path length, $l$, determined by a randomly-selected optical depth \citet{1997A&AS..121...15H}. An abundance file is used as an input, providing the chemical abundance of each species, along with an input SED and files specifying the dust properties, opacities, cross-sections etc while the dust to gas ratio is also specified as an input parameter. Given these, the code self-consistently solves the radiative transfer equations and calculates the gas and dust temperatures, ionisation degree, and the overall emergent SED of the full dust, gas and stars network.

\subsection{The {\sc 3d-pdr} code}\label{3dpdrexplanation}
The {\sc 3d-pdr} code \citep{2012MNRAS.427.2100B} is a three-dimensional astrochemistry code which simulates PDRs of arbitrary density distribution. It solves the chemistry and the thermal balance self-consistently in each computational element of a given cloud and uses the chemical model features of \citet{2006MNRAS.371.1865B}. Like {\sc mocassin}, the code has been used in various extragalactic applications such as modelling molecular line intensities in NGC 4038 \citep{2014MNRAS.443..111B} and neutral carbon mapping \citep{2014MNRAS.440L..81O}. 

{\sc 3d-pdr} uses a ray-tracing scheme based on the HEALPIX \citep{2005ApJ...622..759G} package which calculates properties along a given line of sight. This allows for the quick calculation of a) column densities of species along a particular direction b) the attenuation of the Draine field in the PDR and c) the propagation of the FIR/submm line emission out of the PDR.

The reaction rates within the chemical network are taken from the UMIST 2012 chemical network database in \citet{2013A&A...550A..36M}. Extinction within the cloud is calculated assuming a grain size of $0.1\mu$m, albedo of 0.7 and a mean photon scattering by grains of $g=0.9$. Emission and fine structure lines are calculated using the escape probability method of \citet{1980A&A....91...68D} and non-LTE level populations determined from the collisional rate coefficients explained in \citet{2012MNRAS.427.2100B}. 

Moreover, for the H$_{2}$ and CO photodissociation rates, the code adopts the treatments of \citet{1996A&A...311..690L} and \citet{1988ApJ...334..771V}. To account for the shielding of CI the code uses the treatment of \citet{2000A&A...353..276K} in order to estimate the photoionisation rate of carbon. The rate of molecular hydrogen formation on dust grains is calculated using the treatment of \citet{2004ApJ...604..222C} while the thermally averaged sticking coefficient of hydrogen atoms on dust grains is taken from \citet{1979ApJS...41..555H}. The dust temperature at each point in the density distribution is calculated using the treatment of \citet{1991ApJ...377..192H} to account for the grain heating due to the incident FUV photons. Finally the code also handles varying gas-phase metallicities. The grain surface H$_2$ formation rate of \citet{1977A&A....55..137D} is adopted and scales linearly with metallicity, while the dust to gas ratio also scales linearly with metallicity, taking a standard value of 10$^{-2}$ at solar metallicity, following the prescription by \citet{2011ApJ...737...12L}. The dust and PAH photoelectric heating also scales with metallicity

%%%%%%%%%%%%%%%%%%%%%%%%%%%%%%%%%%%%%%%%%%%%%%%%%%%
%%%%%%%%%%%%%%%%%%%%%%%%%%%%%%%%%%%%%%%%%%%%%%%%%%%
\section{Self-consistent coupling method} \label{couplingmethod}
To self-consistently couple all three of the above codes it is important to use as many outputs from one model as inputs for subsequent models, ensuring consistency within the full simulation. {\as This coupling technique was first attempted in \citet{magdathesis} to model carbon and oxygen emission in nearby galaxies. We describe here how each code is numerically coupled to the other models, with a detailed discussion of the specific input parameters and their values presented in Section \ref{modelparamspace}.} Spherical symmetry is assumed throughout this paper, in all phases of the ISM and for all the simulated 3D clouds. Although {\sc mocassin} and {\sc 3d-pdr} are both fully capable of handling non-uniform densities, this will not be used here. 

\subsection{Coupling {\sc starburst99} to {\sc mocassin}}
A stellar radiation density field, coming from the stellar population within our simulated star forming regions, is created using {\sc sb99}. {\as From this output {\sc sb99} stellar spectrum, the luminosity, temperature and number of ionising photons of the source are calculated; these quantities are then used as input parameters for the 3D photoionisation code {\sc mocassin}. In this way, the radiation field is coupled with the photoionisation in the H\footnotesize II \normalsize region.} 

\subsection{Coupling {\sc mocassin} to {\sc 3d-pdr}}\label{moc3dpdrcoupling}
\citet{2005ApJ...621..328H} showed how dynamical processes, such as gas flows and thermal gas pressure, link the H\footnotesize II \normalsize and PDR regions, which are simulated here by {\sc mocassin} and {\sc 3d-pdr}, respectively. The physical properties of the PDR are a consequence of the transport of gas, dust and radiation through the ionised region, while the converse is also true. Other than in the case of a very fortuitous choice of initial conditions, simply matching boundary conditions between quantities such as temperature and density, of the two regions, can lead to discontinuities in the thermal and dynamical pressure across the two ISM phases. Only by physically coupling the two regions is it possible to get an accurate representation of the multi-phase ISM; this modelling philosophy is at the heart of this work as we aim to self-consistently calculate the temperature, ionisation, and density at the face of PDR regions.

{\as The output of the {\sc mocassin} code is the SED of the ionised gas, dust and stars emerging from the H\footnotesize II \normalsize region, along with the flux in the most important far infrared fine structure emission lines from the ionised gas, such as \cii.} We calculate the strength of the radiation field, G$_{0}$, at the ionisation front between the H\footnotesize II \normalsize and neutral gas regions by integrating the {\sc mocassin} SED in the far-UV range between 912 to 2400\AA, which is the classical Draine field definition \citep{1978ApJS...36..595D}. This value of G$_{0}$ is used as an input into {\sc 3d-pdr}. We also need to ensure that {\sc mocassin} only simulates the gas up to the edge of the ionised region and does not leak over into the neutral regions (which {\sc 3d-pdr} will simulate). To this end, we first calculate the outer radius of the ionised cloud simply by running {\sc mocassin} to a very large radius, and then inspecting at which radius the ionised Hydrogen abundance is less than 10\%, ensuring the end of the ionised region has been reached; this is taken to be the outer radius of the ionised part of the star forming region. The inner radius of the PDR is equal to the outer radius of the ionised region, ensuring that the PDR is adjacent to the ionised region (see Section \ref{modellingmocassin} for further details).

We further link the H\footnotesize II \normalsize and PDR regions by assuming constant total pressure at the interface, ensuring the temperature and density of the gas and dust between the two regions are self-consistently calculated. Constant pressure is an approximation to the actual flow which has been assumed by previous authors such as \citet{1994ApJ...423..223C}, \citet{2005ApJS..161...65A} and \citet{2013RMxAA..49..137F}. Pressure originates from the stellar continuum and internally generated light, gas pressure, and sometimes from turbulence, ram, and magnetic pressures when appropriate \citep{1991ApJ...374..580B, 2005ApJ...621..328H}. We are only interested in terms which change across the boundary and hence focus on the gas and dust terms as all others remain constant. The pressure terms of interest originate from internal radiation, from the gas and dust, and the thermodynamic gas pressure. We assume that the gas and dust emission is optically thick as we are only interested in solving the equation at the boundary between the two regions. Therefore, we set up an equation of pressure balance as follows:
% pressure equation of state
\begin{dmath}
\centering
n_{\mbox{\scriptsize{H\tiny II \normalsize}}}k_{\mbox{\scriptsize{b}}}T_{\mbox{\scriptsize{H\tiny II \normalsize}}}^{\mbox{\scriptsize{gas}}} + \frac{\epsilon_{\mbox{\scriptsize{gas}}}^{\mbox{\scriptsize{H\tiny II \normalsize}}}\sigma_{\mbox{\scriptsize{b}}} T_{\mbox{\scriptsize{H\tiny II \normalsize}}}^{4,\mbox{\scriptsize{gas}}}}{c} + \frac{\epsilon_{\mbox{\scriptsize{dust}}}^{\mbox{\scriptsize{H\tiny II \normalsize}}}\sigma_{\mbox{\scriptsize{b}}} T_{\mbox{\scriptsize{H\tiny II \normalsize}}}^{4,\mbox{\scriptsize{dust}}}}{c} =
n_{\mbox{\scriptsize{PDR}}}k_{\mbox{\scriptsize{b}}}T_{\mbox{\scriptsize{PDR}}}^{\mbox{\scriptsize{gas}}} + \frac{\epsilon_{\mbox{\scriptsize{gas}}}^{\mbox{\scriptsize{PDR}}} \sigma_{\mbox{\scriptsize{b}}} T_{\mbox{\scriptsize{PDR}}}^{4,\mbox{\scriptsize{gas}}}}{c} + \frac{\epsilon_{\mbox{\scriptsize{dust}}}^{\mbox{\scriptsize{PDR}}} \sigma_{\mbox{\scriptsize{b}}} T_{\mbox{\scriptsize{PDR}}}^{4,\mbox{\scriptsize{dust}}}}{c}
\label{pressureequation}
\end{dmath}
where $n_{\mbox{\scriptsize{H\tiny II \normalsize}}}$ and $n_{\mbox{\scriptsize{PDR}}}$ are the electron number densities in the H\footnotesize II \normalsize and PDR region, respectively, $k_{\mbox{\scriptsize{b}}}$ is the Boltzmann Constant, $\sigma_{\mbox{\scriptsize{b}}}$ is the Stefan Boltzmann constant, while $\epsilon_{\mbox{\scriptsize{gas}}}$ and $\epsilon_{\mbox{\scriptsize{dust}}}$ are the emissivities of the gas and dust species. $T_{\mbox{\scriptsize{H\tiny II \normalsize}}}^{\mbox{\scriptsize{gas}}}$ and $T_{\mbox{\scriptsize{H\tiny II \normalsize}}}^{\mbox{\scriptsize{dust}}}$ are the gas and dust temperatures at the edge of the H\footnotesize II \normalsize region, as calculated from the {\sc mocassin} output. Since $n_{\mbox{\scriptsize{H\tiny II \normalsize}}}$ is one of the input parameters of our code, the above equation needs to be solved for for $n_{\mbox{\scriptsize{PDR}}}$, $T_{\mbox{\scriptsize{PDR}}}^{\mbox{\scriptsize{gas}}}$ and $T_{\mbox{\scriptsize{PDR}}}^{\mbox{\scriptsize{dust}}}$. 

Motivated by dust temperature continuity across the two regions obtained in other self-consistent calculations such as {\sc cloudy} \citep{2013RMxAA..49..137F} and {\sc Torus-3dpdr} \citep{2015MNRAS.454.2828B} we set
\begin{dmath}
\centering
T_{\mbox{\scriptsize{H\tiny II \normalsize}}}^{\mbox{\scriptsize{dust}}}=  T_{\mbox{\scriptsize{PDR}}}^{\mbox{\scriptsize{dust}}}.
\end{dmath}
We further assume the same dust species in both regions, leading to the cancellation of the terms describing radiation pressure from dust emission in Equation \ref{pressureequation}. The equation of pressure balance solely for the gas remains, such that:
\begin{dmath}
\centering
n_{\mbox{\scriptsize{H\tiny II \normalsize}}}k_{\mbox{\scriptsize{b}}}T_{\mbox{\scriptsize{H\tiny II \normalsize}}}^{\mbox{\scriptsize{gas}}} + \frac{\epsilon_{\mbox{\scriptsize{gas}}}^{\mbox{\scriptsize{H\tiny II \normalsize}}}\sigma_{\mbox{\scriptsize{b}}} T_{\mbox{\scriptsize{H\tiny II \normalsize}}}^{4,\mbox{\scriptsize{gas}}}}{c} = 
n_{\mbox{\scriptsize{PDR}}}k_{\mbox{\scriptsize{b}}}T_{\mbox{\scriptsize{PDR}}}^{\mbox{\scriptsize{gas}}} + \frac{\epsilon_{\mbox{\scriptsize{gas}}}^{\mbox{\scriptsize{PDR}}} \sigma_{\mbox{\scriptsize{b}}} T_{\mbox{\scriptsize{PDR}}}^{4,\mbox{\scriptsize{gas}}}}{c}.
\end{dmath}
Due to the low emissivity of gas we make the approximation that the radiation pressure caused by photons emitted from the gas can be ignored and so this term is set to zero. Therefore, to set the conditions in the PDR, caused by the ionised region, we are left to solve: 
\begin{dmath}
\centering
n_{\mbox{\scriptsize{H\tiny II \normalsize}}}T_{\mbox{\scriptsize{H\tiny II \normalsize}}}^{\mbox{\scriptsize{gas}}}  = n_{\mbox{\scriptsize{PDR}}}T_{\mbox{\scriptsize{PDR}}}^{\mbox{\scriptsize{gas}}}.
\label{newton} 
\end{dmath}
The temperature at the surface of the PDR is dependent on its hydrogen number density, $n_{\mbox{\scriptsize{PDR}}}$, therefore the above equation can be solved using a Newton-Raphson numerical method to provide the value of the hydrogen number density of the PDR given the conditions in the H\footnotesize II \normalsize region\footnote{We iteratively solve Equation \ref{newton} up to a 1\% accuracy level.}, ensuring self-consistency. 

Overall our method leads to dust temperature continuity between the two regions, and also ensures gas pressure equilibrium. The resulting temperature and density profiles of the gas and dust across the two regions are consistent with profiles obtained in {\sc cloudy} \citep{2013RMxAA..49..137F} and {\sc Torus-3dpdr} \citep{2015MNRAS.454.2828B}. 
\section{Model Parameter Space} \label{modelparamspace}
The main purpose of this work is to provide a prescription to calculate the fraction of the total integrated \cii\ emission of a galaxy emanating from the molecular phase of the ISM, \fmol, from typical extragalactic observables such as stellar mass, SSFR and metallicity.  {\as Therefore, we are interested in running our self-consistent modelling interface over an input parameter space corresponding to meaningful observables on galaxy wide scales. To this end, parameters which do not correspond to galactic observables shall be kept constant and typical values shall be used and taken from the literature. Before presenting the results of the modelling, we describe in this section the parameters chosen for the different codes, and the final seven input parameters that are required to run the full coupled model.  }

\subsection{Stellar population parameters in {\sc starburst99}}
\begin{figure}
  \centering
    \includegraphics[scale=0.55]{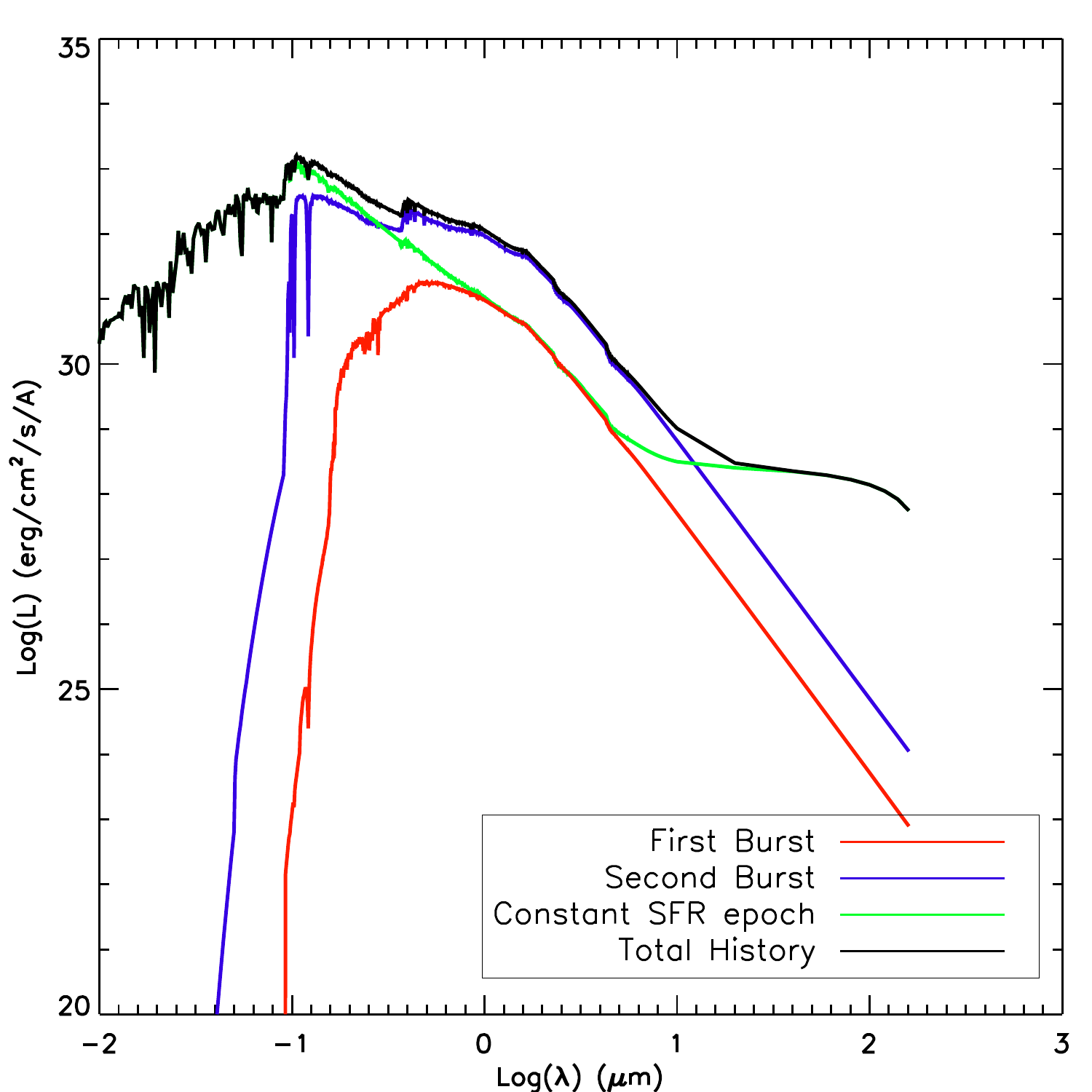}
\caption{Example stellar SEDs from the population of stars created throughout the whole star formation history of our clouds. The instantaneous burst and constant star formation rate epoch both compete to dominate the UV part of the spectrum.}
\label{sed1} 
\end{figure}
{\as We simulate the total stellar SED as originating from a single ionising source (even though it physically originates from multiple sources) for simplicity and ease. \citet{2013MNRAS.431.2493K} have compared the outputs of {\sc mocassin} when the ionising flux is produced by a single source or by 100 sources distributed uniformly within a sphere of 0.2 kpc. They find the results to vary only at small radii; outside of the inner radius, the gas effectively sees a point source. As we are here integrating far out of the cloud, this simplifying assumption will not affect our results.  

The specific SED produced by {\sc sb99} depends on assumptions made regarding the star formation history, IMF, and metallicity of the stellar population. The code allows input stellar metallicities of 0.02, 0.2, 1 or 2.5 Z$_{\odot}$, and either periods of constant star formation or instantaneous bursts. Although the stellar metallicity is not a variable parameter in this work, it is undoubtedly correlated with the gas-phase metallicity, which will be a variable parameter in our models. Within these constraints, we build the star formation histories of our model galaxies as follows: (1) a first instantaneous burst of star formation at approximately the Hubble time with the total stellar mass produced in this burst given as an input parameter, and a stellar metallicity of 0.02Z$_{\odot}$ as the metal content of the early universe is negligible, (2) a period of quiescence followed by a secondary burst, with the age of this second burst another input parameter, and (3) a period of constant star formation until the present day, with this rate of star formation a third input parameter. Due to metal enrichment of the gas from previous supernovae, we set the stellar metallicity of phases (2) and (3) to be the available input parameter greater than the input gas-phase metallicity, e.g for a gas-phase metallicity of 0.65Z$_{\odot}$ we would set the stellar metallicity to 1Z$_{\odot}$. The mass lost to supernovae and stellar winds in the initial burst is calculated and fed back into the secondary burst, ergo the input stellar mass parameter for phase (2), ensuring that the total stellar mass of the star forming region (one of the input parameters) is successfully produced by the present day. We do this in keeping with the current paradigm that star formation is regulated by outflows from stellar winds and supernovae \citep{2011MNRAS.415...11D, 2012MNRAS.421...98D, 2016EAS....75..359W}.

{\sc sb99} only allows the use of a piecewise power law IMF and so we build one which closely matches the IMF of \citet{2003PASP..115..763C}. We use exponents of 2.3 and 1.3 for IMF boundaries of 1.0 $<$ M$_{\ast}$/M$_{\odot}$ $<$ 100.0 and 0.1 $<$ M$_{\ast}$/M$_{\odot}$ $<$ 1.0 respectively. The latter boundary exponent choice is equivalent to the IMF of \citet{2001MNRAS.322..231K}, which is approximately equal to that of \citet{2003PASP..115..763C} as noted by \citet{2014ApJS..214...15S}. When running {\sc sb99}, we use the Padova stellar evolution tracks, detailed in \citet{1994A&AS..105...29F}, with thermally pulsating AGB stars and Pauldrach and Hiller model atmospheres. Figure \ref{sed1} shows an example of a stellar SED produced by {\sc sb99} for such a star formation history, with the contribution of the three different phases also shown separately. } %; our input IMF piecewise power law is as close to a \citet{2003PASP..115..763C} IMF as {\sc sb99} allows

\subsection{Ionised region parameters in {\sc mocassin}}\label{modellingmocassin}
The {\sc moccasin} simulations employ 3D spherically symmetric geometry, with the ionising source at the centre of the clouds. The inner radius of the ionised gas region is set to as close to zero as computationally possible. To determine the outer radius of the ionised gas region, we run {\sc mocassin} up to a large outer radius and calculate the radius at which the ionised hydrogen fraction drops below ten percent (this is the effective Str$\ddot{\mbox{o}}$mgren sphere radius). We then take this calculated radius value and re-run {\sc mocassin} but now setting the outer radius to this calculated radius. Hydrodynamical effects such as turbulence, shocks and magnetic fields are ignored in our simulations. 

The input stellar spectrum, the source luminosity and temperature are all taken from {\sc sb99}. {\sc mocassin} calculates from them the number of ionising photons per second, Q$_{\mbox{\small phot}}$\normalsize. The 3D grid used for the simulations has $15\times15\times15$ resolution elements so as not to be too computationally expensive, while also ensuring no loss of detail via blending across cells. The two input parameters we are free to vary are the electron number density of the H\footnotesize II \normalsize region, which takes typical values from $10^{1.5}$ to $10^{3}$ cm$^{-3}$, and the gas-phase metallicity which we vary between 0.2-1.1Z$_{\odot}$. The metallicity determines the dust to gas ratio input as, to ensure consistency with {\sc 3d-pdr}, we use the prescription by \citet{2011ApJ...737...12L} (as discussed in \ref{3dpdrexplanation}). We use different grain properties for H\footnotesize II \normalsize and PDR regions due the the different physical conditions found in these regions. Within {\sc mocassin} we use the standard silicate dust properties detailed in \citet{1984ApJ...285...89D}. For the PDR region we use a mixture of silicates + PAHS + graphite, due to the higher column densities, with graphite grains being the dominant dust species. 

\begin{table}
\centering
 \caption{Gas-phase elemental abundances used in {\sc mocassin} and {\sc 3d-pdr}, relative to total hydrogen number density, at solar metallicity. All these elements, except Hydrogen and Helium which are primordial in origin, scale linearly with metallicity.}
 \begin{tabular}{@{}ll@{}}
 \hline
  Species     &  Gas-phase abundance   \\
\hline
He/H &    0.1 \\
O/H &  4.9 $\times$ 10$^{-4}$ \\
N/H &  6.9 $\times$ 10$^{-5}$ \\
Ne/H & 1.1 $\times$ 10$^{-4}$ \\
S/H &  8.1 $\times$ 10$^{-6}$ \\
Ar/H & 1.9 $\times$ 10$^{-6}$ \\
C/H &  3.6 $\times$ 10$^{-4}$ \\
Si/H &  4.8 $\times$ 10$^{-6}$ \\
Mg/H &  4.0 $\times$ 10$^{-5}$ \\
Fe/H & 3.6 $\times$ 10$^{-6}$ \\
\hline
\end{tabular}
\label{abundances}
\end{table}

%section 4.3
\subsection{Photodissociation region parameters in {\sc 3d-pdr}}
For {\sc 3d-pdr}, we consider a spherically symmetric shell of uniform density neutral and molecular gas, surrounding the ionised region. The inner radius of the PDR region is therefore the outer radius of the ionised region as calculated with {\sc mocassin}, and the outer radius (and corresponding $A_v$) is set by the dust mass fraction, M$_{dust}$/M$_{\ast}$; we integrate out to a radius that is set by the dust mass budget available.  We define the molecular region as the region where more than 1\% of hydrogen is in molecular form, marking the beginning of the CO-dark phase. Geometrical dilution effects of the UV field are taken into account to obtain accurate 3D results. 
\begin{table*}
\caption{Variable input parameters used in the fully coupled multi-phase code.}
 \begin{tabular}{@{}lllll@{}}
 \hline
  Input parameter     &  \multicolumn{1}{c}{Minimum value}   &  \multicolumn{1}{c}{Maximum value}   & \multicolumn{1}{c}{Number of Variations} &  \multicolumn{1}{c}{Model}  \\
\hline
Gas-phase metallicity & 0.20Z$_{\odot}$ & 1.1Z$_{\odot}$ & \multicolumn{1}{c}{4} & \multicolumn{1}{c}{{\sc mocassin}, {\sc 3d-pdr}, {\sc sb99}}  \\
Stellar mass of the cloud & 10$^{2}$ M$_{\odot}$ & 10$^{4}$ M$_{\odot}$ & \multicolumn{1}{c}{3} & \multicolumn{1}{c}{{\sc sb99}}  \\
Stellar population age & 10$^{2}$Myr & 10$^{3.0}$Myr & \multicolumn{1}{c}{3} & \multicolumn{1}{c}{{\sc sb99}} \\
H\scriptsize II \small region electron number density & 10$^{1.5}$ cm$^{-3}$ & 10$^{3.0}$ cm$^{-3}$  & \multicolumn{1}{c}{4} & \multicolumn{1}{c}{{\sc mocassin}}\\
Cosmic ray ionisation rate & 10$^{-17}$ s$^{-1}$  &  10$^{-14}$ s$^{-1}$ & \multicolumn{1}{c}{4} & \multicolumn{1}{c}{{\sc 3d-pdr}}\\
Dust mass fraction & 10$^{-4}$ & 10$^{-2}$  &  \multicolumn{1}{c}{5} & \multicolumn{1}{c}{{\sc 3d-pdr}}\\
Specific star formation rate & 10$^{-11.5}$ yr$^{-1}$ & 10$^{-9.5}$ yr$^{-1}$ & \multicolumn{1}{c}{3} & \multicolumn{1}{c}{{\sc sb99}} \\
\hline
\end{tabular}
\label{param_space}
\end{table*}
We assume a standard turbulent velocity of 1.5 kms$^{-1}$, while the hydrogen number density is self-consistently calculated\footnote{This turbulence contributes a negligible amount to the total pressure between the two regions, and hence is excluded in Equation \ref{pressureequation}} (see Section \ref{moc3dpdrcoupling}). Therefore, the two input parameters we are free to vary for the PDR regions are the cosmic ray ionisation rate and the dust mass fraction.  The gas-phase metallicity is taken to be the same as the value selected for the ionised region.  Abundances of all metals scale linearly with metallicity, and Table \ref{abundances} summarises the initial chemical abundances used in both {\sc mocassin} and in {\sc 3d-pdr} at solar metallicity. We use identical chemical abundances between the codes to maintain self-consistency and take the abundances at solar metallicity from \citet{2012A&A...548A..20C}. 

%section 4.4
\subsection{Summary of input parameters}
Our objective is to provide a prescription for variations of \fmol\ in galaxy-integrated observations that can be applied to unresolved galaxy-wide observations.  Therefore, we use as input parameters quantities that are motivated by galaxy-wide observations, where possible. Choices for all the input parameters of our multi-phase ISM code are justified here and summarised in Table \ref{param_space}. 
\begin{itemize}
 \item \textit{Stellar mass} - We let the stellar mass of our simulated star-forming regions vary from 10$^{2}$ to 10$^{4}$ M$_{\odot}$. These values are typical of star-forming regions within the Milky Way \citep{2010ApJ...713..871W}. 
  
 \item \textit{Age of the secondary burst} - Since our choice of star formation histories is meant to reproduce a broad range of possible integrated population ages, we choose to probe a wide range for the time since the secondary burst, spanning over 1.5 dex from 10$^{2}$ - 10$^{3.5}$ Myr.  
 
 \item \textit{Specific star formation rate} - Deep multi-wavelength extragalactic surveys have revealed a tight correlation between SFR and stellar mass for star-forming galaxies \citep[e.g.][]{2007ApJ...660L..43N,2007A&A...468...33E,2007ApJ...670..156D}. This correlation is well-established in the local universe and up to $z\sim3$ \citep[e.g.][]{2004MNRAS.351.1151B,2010ApJ...721..193P,2012ApJ...754...25R}.  We choose SSFRs in the range of $10^{-11.5}-10^{-9.5}$ yr$^{-1}$, typical of main sequence galaxies in the local universe, using the stellar mass of the star-forming regions for normalisation.
 
 \item \textit{Gas-phase metallicity} - We use the mass metallicity relation from \citet{2004ApJ...613..898T} to guide this choice, and adopt a metallicity range of 0.2 - 1.1Z$_{\odot}$ to reproduce conditions in local universe star-forming galaxies with \mstar$>10^9$\msun. 
 
 \item \textit{Electron number density of the H\footnotesize II \normalsize region} - We choose to vary the hydrogen number density between 10$^{1.5}$ and  10$^{3.0}$ cm$^{-3}$ based on the values calculated by \citet{2009A&A...507.1327H} for extragalactic H\footnotesize II \normalsize regions. 
  
 \item \textit{Cosmic Ray Ionisation Rate} - In the local universe, this is known to be roughly 10$^{-17}$ s$^{-1}$ - 10$^{-16}$ s$^{-1}$ \citep{PoS(ICRC2015)318, 2006PNAS..10312269D}, but values can be larger by up to three orders of magnitude in galaxies with very large SFRs such as local ULIRGs and high-redshift star-forming galaxies \citep{2010ApJ...720..226P}. We therefore explore a range of cosmic ray ionisation rates ranging from 10$^{-17}$ to 10$^{-14}$ s$^{-1}$ to allow us to also explore conditions typical of $z\sim2$ galaxies.
 
  \item \textit{Dust Mass Fraction} - We run models with the dust mass fraction (M$_{dust}$/M$_{\ast}$) varying between 10$^{-4}$ and 10$^{-2}$ based on the scaling relation between dust mass fraction and stellar mass derived from the galaxies in the Herschel Reference Survey \citep{2010PASP..122..261B}. 
  
\end{itemize}

%%%%%%%%%%%%%%%%%%%%%%%%%%%%%%%%%%%%%%%%%%%%%%%%%%%
%%%%%%%%%%%%%%%%%%%%%%%%%%%%%%%%%%%%%%%%%%%%%%%%%%%
\section{Numerical Results} \label{numericalresults}

%\begin{figure*}
 %\centering
  % \includegraphics[scale=1.0]{with_post_processing_scaling_relations_results.png}
%\caption{Scaling relations for all the simulated clouds which reached thermal equilibrium. Due to the large number of data points we provide a colour coding which indicates the one dimensional normalised number density, within each bin of our parameter space. Red points represent areas within each bin having the highest number of data points. In each panel, the dotted line is a quadratic fit to the data. }
%\label{all_scaling_relations_processed}
%\end{figure*}

To cover the full parameter space summarised in Table \ref{param_space}, we simulated a total of 8640 individual star forming regions. Of those, 8016 clouds fully converged to a solution achieving thermal equilibrium. The main quantity of interest in the context of this study is \fmol, the fraction of the total \cii\ emission originating from the molecular regions. A quantitative analysis of the data requires a multi-dimensional hierarchical Bayesian inference method, which will be performed in Section \ref{statsmethod}. For now, we qualitatively investigate the dependence of \fmol\ on the input parameters of the model and explain the physics behind the trends which emerge. 

%For now, we qualitatively explain the physics behind the trends observed in Fig. \ref{all_scaling_relations_processed} and the dependence of \fmol\ on the input parameters of the model. 

% In Fig. \ref{all_scaling_relations_processed}, we plot scaling relations for \fmol\ as a function of the input parameters of the coupled model.   It can be seen that no one input parameter is solely responsible for variations in \fmol. The strongest trend is seen with dust mass fraction. This is expected, as this parameter determines the size of each cloud, and therefore how deep into the PDR regions we perform the integration. The higher the dust mass fraction, the further we integrate into the PDR regions leading to an increase in \fmol. However, the dispersion is still very high indicating that other parameters do play a role. 

\subsection{Stellar mass and star formation rate}\label{sfr_explanation}
\begin{figure*}
\includegraphics[scale=0.5]{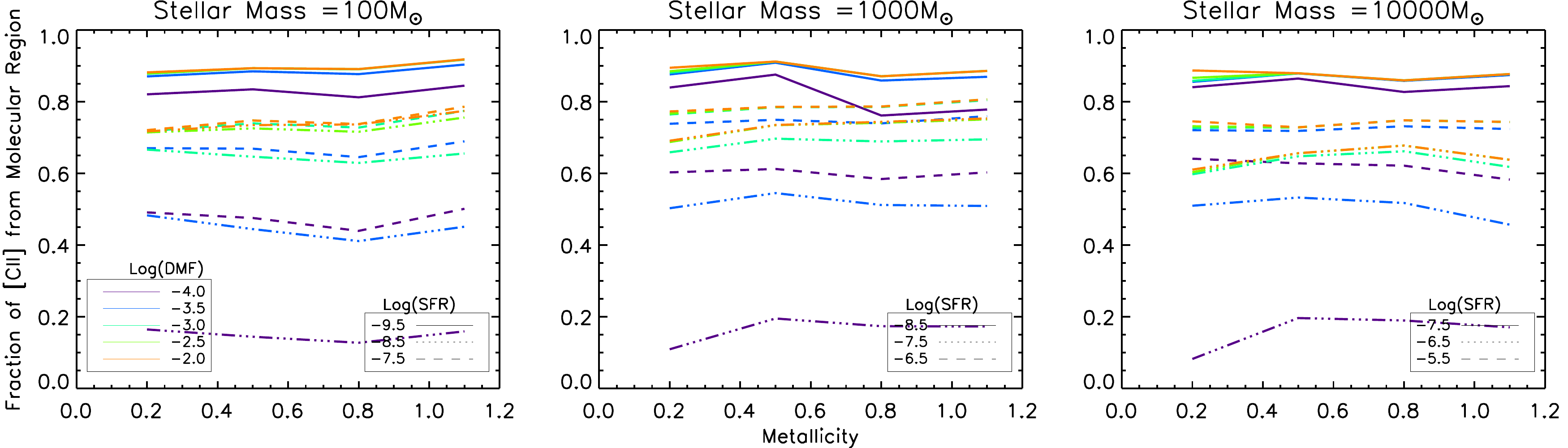}
\caption{Variations of \fmol\ as a function of metallicity for three different stellar mass bins. In each panel the relation is shown for different values of the dust mass fraction (different colours) and SFR (different linestyles). The hydrogen number density, cosmic ray ionisation rate and age of the secondary burst of star formation are kept constant.}
\label{mass_analysis}
    \includegraphics[scale=0.5]{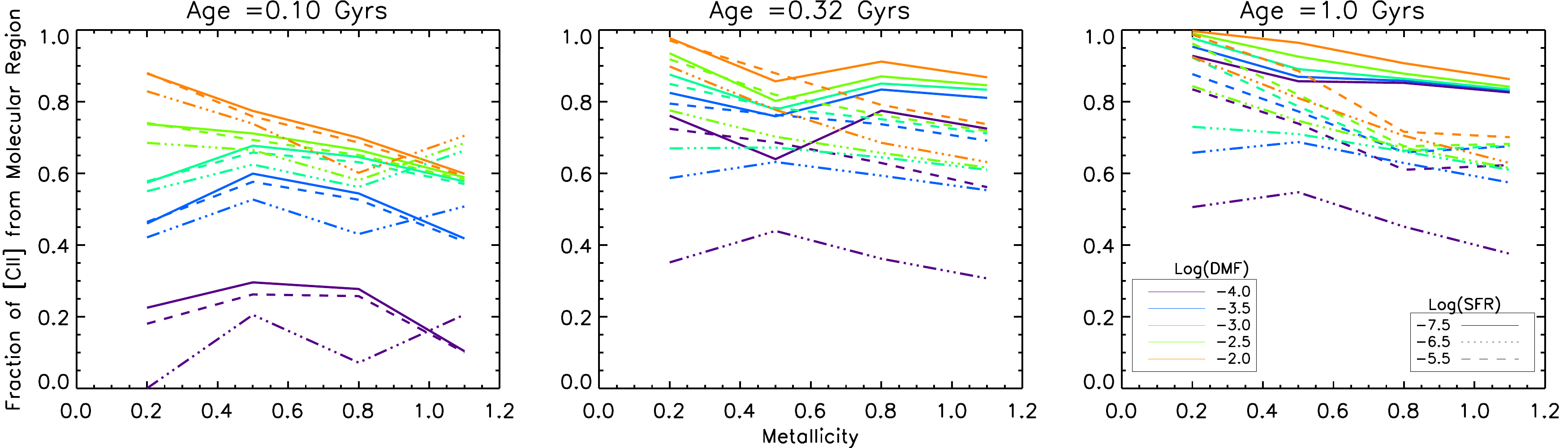}
\caption{Variations of \fmol\ as a function of metallicity for three different ages for the secondary burst of star formation. In each panel the relation is shown for different values of the dust mass fraction (different colours) and SFR (different linestyles). The hydrogen number density, cosmic ray ionisation rate and stellar mass are kept constant.}
\label{age_analysis}
\end{figure*}

We first show in Fig. \ref{mass_analysis} how \fmol\ varies as a function of metallicity for three different values of stellar mass.  At fixed mass and metallicity, the figure also shows the impact of a varying the dust mass fraction and SFR.  In this example, the hydrogen number density, the cosmic ray ionisation rate and the age of the secondary burst of star formation are kept fixed.  The figure shows that \fmol\ does not vary significantly with either stellar mass or metallicity. This is as expected because, as the stellar mass decreases, the number of ionising photons also decreases, reducing the overall size of the cloud. However, the relative sizes and densities of the H\footnotesize II \normalsize to PDR regions will not change. By scaling down the stellar mass at the centre of each cloud we have simply scaled down the size of the cloud while maintaining the same physical structure throughout each cloud. 

Figure \ref{mass_analysis} also shows that at fixed stellar mass, metallicity and dust mass fraction, an increase in SFR corresponds to a decrease in \fmol. This is because, at fixed dust mass, an increase in the star formation rate leads to an increase in the radius of the H\footnotesize II \normalsize regions as more photoionising UV photons are available. More \cii\ will therefore arise from the ionised regions versus the molecular regions, and hence \fmol\ decreases.

\subsection{Age of secondary burst}
The variations caused by the age of the secondary burst are of a similar nature to star formation rate. Star formation histories which involve a younger secondary burst provide more photoionising UV photons. Hence, the younger the age of the secondary burst, the more \cii\ will emerge from the H\footnotesize II \normalsize regions, as the Str\"omgren sphere radius increases. This can be seen in Fig. \ref{age_analysis}, with the mean value of \fmol\ increasing as the age of the secondary burst increases from 0.1 to 0.32 to 1.0 Gyr. This effect is less pronounced than that caused by variations in star formation rate, because the majority of UV photons are produced by the low level star formation happening at the present time rather than by the secondary burst (see Fig. \ref{sed1}). Interestingly though, the time since the last burst of star formation nonetheless has a detectable effect on \fmol\ variations.

\subsection{Gas-phase metallicity}
Metallicity is responsible for variations in a more complex manner, with two main effects competing for dominance. To investigate these two processes we refer to Figs. \ref{mass_analysis} and \ref{age_analysis}. 

One could naively expect that by decreasing the amount of metals available throughout the whole system, that the abundance of carbon in the ionised, neutral and molecular regions would decrease in equal measure and hence no variations of \fmol\ should be seen because of metallicity (a similar argument to the lack of variations caused by stellar mass). However, within the ionised regions, the cooling rate is a function of metallicity; a decrease in metallicity leads to a lower cooling rate and therefore an increase in the size of the Str\"omgren sphere. Hence, from this first effect (the Cooling Rate effect, hereafter), we can expect that by decreasing metallicity there will follow a decrease of \fmol. 

However a second, more dominant effect, is the well-known photodissociation of CO into ionised carbon (The Photodissociation effect, hereafter). In low metallicity environments, FUV radiation penetrates further into the clouds leading to an enhanced abundance of ionised carbon in the molecular regions. The Photodissociation effect therefore has the opposite effect of increasing \fmol\ as metallicity decreases. 

In different parts of parameter space, the Cooling Rate and the Photodissociation effects cancel each other out, leading to negligible variations of \fmol\ as a function of metallicity, as seen in Figs. \ref{mass_analysis} and Fig. \ref{age_analysis}.  Under other circumstances, the Cooling Rate or the Photodissociation effect dominates, leading to positive or negative slopes in the \fmol-Z relation, respectively. 

\subsection{Electron number density of the ionised region.}\label{electron_density_explanation}
\begin{figure*}
  \centering
    \includegraphics[scale=0.4]{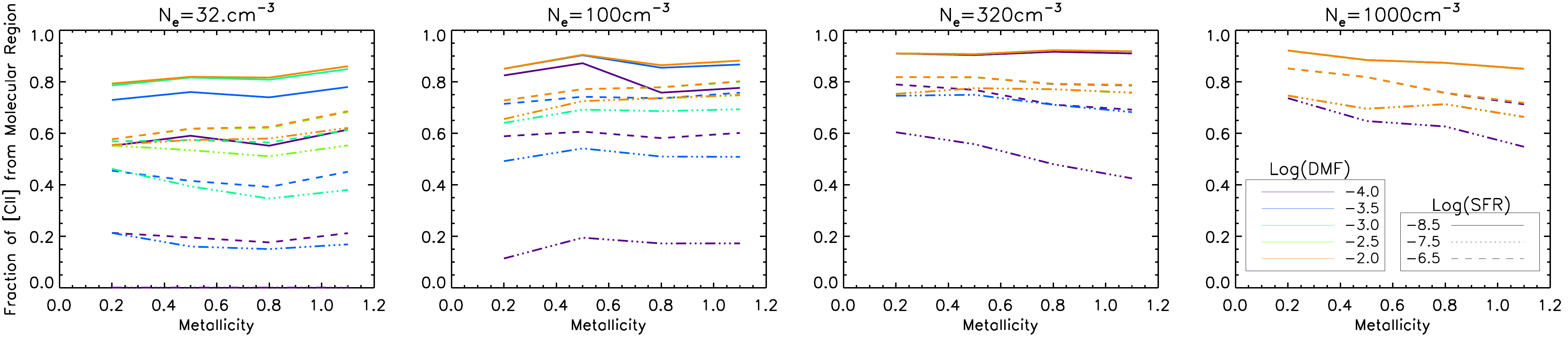}
\caption{Variations of \fmol\ as a function of metallicity for four different density bins. In each panel the relation is shown for different values of the dust mass fraction (different colours) and SFR (different linestyles). The stellar mass, cosmic ray ionisation rate and age of the secondary burst of star formation are kept constant.}
\label{density_analysis}
\end{figure*}

An increase in the density of the ionised region ($n_e$) leads to no change on the \cii\ emission from this region as we've already reached the critical density for collisions with electrons, of $\sim$50cm$^{-3}$ \citep{2012ApJS..203...13G}, which dominate in the H\footnotesize II \normalsize region. However, due to equilibrium, this leads to an increase in the density of the PDR allowing for an increase in the \cii\ emission from the molecular region, and correspondingly, an increase of \fmol. This continues until we reach the critical density for collisions with hydrogen in the PDR, of $\sim$10$^{3.5}$cm$^{-3}$ \citep{2012ApJS..203...13G}, which dominate in the neutral ISM phases, at which point \fmol\ remains roughly constant. We can see these variations caused by the electron number density in the ionised regions in Fig. \ref{density_analysis}. We keep the cosmic ray ionisation rate, stellar mass and age of the secondary burst constant and see that increasing $n_e$ leads to an increase of \fmol, in all dust mass fraction and star formation rate bins.  

The variations caused by density are linked with those of metallicity. In lower density environments, which have larger ionised regions, the metallicity variations are either flat or have a positive gradient implying that the Cooling Rate effect, detailed above, is more dominant. However,  as density increases, the slope of the \fmol-Z relation changes as the Photodissociation effect begins to dominate. Understanding quantitatively how, and when, these effects dominate follows in Section \ref{statsmethod}. 

\begin{figure}
  \centering
    \includegraphics[scale=0.5]{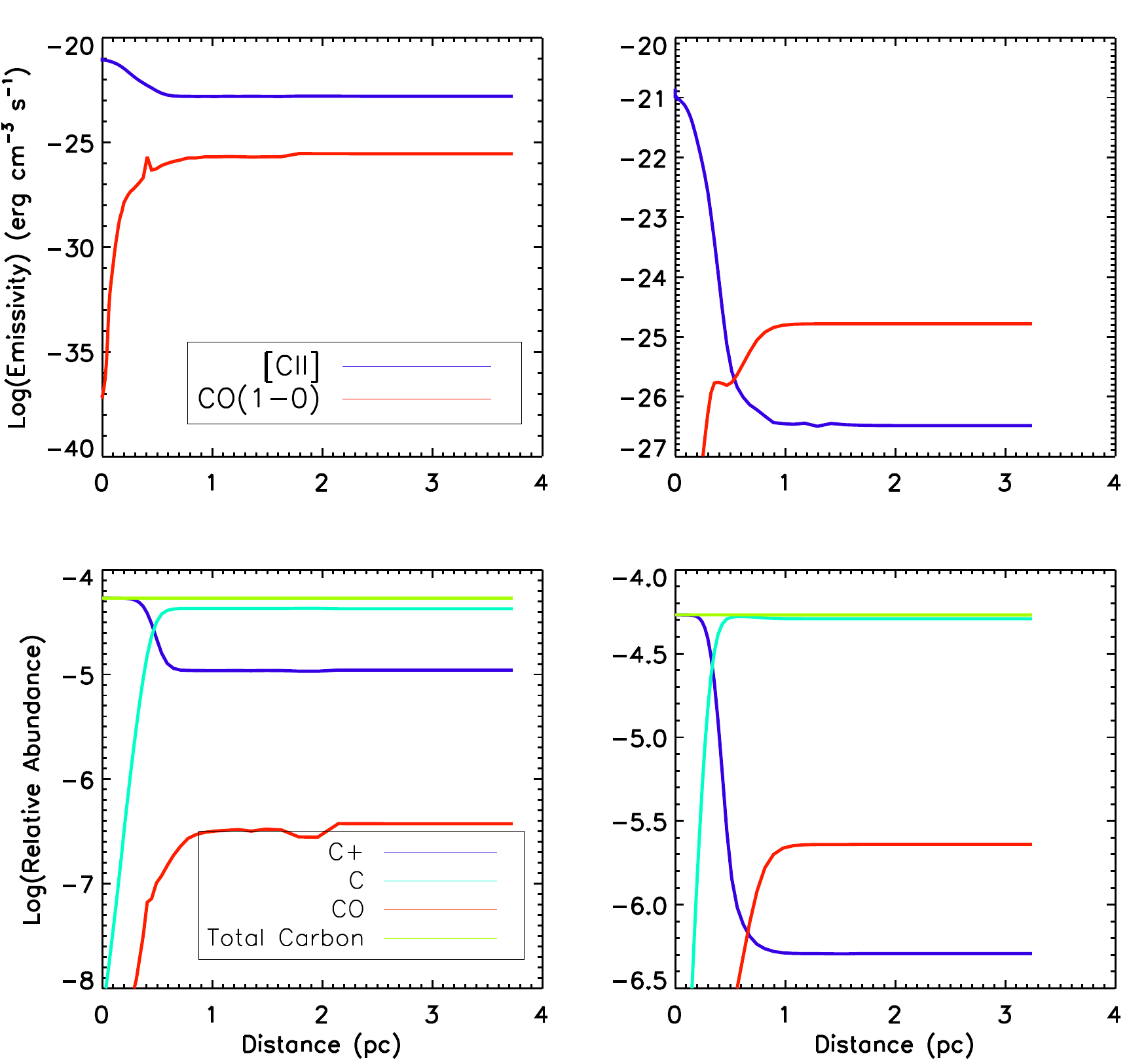}
\caption{\as Examples of a warm (left) and cold (right) clouds. The top row shows for each example cloud how the emissivity of the different carbon phases varies with radius in the cloud, while the bottom two shows the relative abundance of these species. In warm clouds, even within PDRs, the emission of ionised carbon always remains larger than that of CO(1-0), while in cold clouds the CO(1-0) molecular phase dominates.}
\label{profiles}
\end{figure}

\subsection{Cosmic ray ionisation rate}
The cosmic ray ionisation rate input parameter is used only in the molecular and neutral regions (i.e the PDR) where the heating function increases with the cosmic ray ionisation rate. This creates higher temperatures in the PDR regions, which leads an increase in the [C\small II\normalsize] PDR emission, as this line is a major coolant of the gas, leading to an increase in \fmol. %Work done by \citet{2015ApJ...803...37B} has shown that \cii\ starts to dramatically increase for cosmic ray values greater than 10$^{-15}$ s$^{-1}$, explaining the trend in Fig. \ref{all_scaling_relations_processed} when going from lower to higher cosmic ray ionisation rates. 

\subsection{Dust Mass Fraction}
As discussed above, the dust mass fraction effectively controls the total size of our clouds and determines how far into the PDR were integrate up to.  In Fig. \ref{profiles} we show for two different example clouds how the emissivity and carbonaceous species abundances relative to Hydrogen vary as a function of the radius in the neutral and molecular regions. Under certain conditions 'warm clouds' are simulated such that the temperature of the PDR never reaches 10K, the freeze-out temperature of hydrogen onto dust grains. This occurs for clouds with low PDR densities, high cosmic rays ionisation rates and high gas-phase metallicities. In these warm clouds the emissivity of \cii\ always remains larger than that of CO (top left panel). This is because of the warmer conditions, and because the relative abundance of ionised carbon is also always larger than its molecular counterpart (bottom left panel). Therefore, by increasing the dust mass fraction, we are able to retrieve more \cii\ the deeper we integrate, and so \fmol\ increases with the dust mass fraction.

However in 'colder' conditions, where the temperature of the PDR eventually reaches 10K, the emissivity of ionised carbon decreases deep into the molecular regions, where molecular emission begins to dominate (top right panel). Therefore, increasing the dust mass fraction (i.e integrating further into the cloud) does not affect the relative emission of [C\small II\normalsize]. This can also been seen in the abundance profile (bottom right panel), where the molecular carbon abundance now dominates deep into the clouds. Hence, in cases such as these, increasing the dust mass fraction will only increase \fmol\ up to a point before no more \cii\ is obtained and increasing dust mass fraction makes no difference. 

We find that 6513, out of our 8016, clouds harbour these 'warmer' conditions with \cii\ emission dominating over CO(1-0). The emergence of these two groups of clouds has been physically observed in the Galactic Plane \citep{2014A&A...561A.122L} where 557/1804 clouds observed there where detected in \cii\ with no CO. Our fraction of warm clouds to cold clouds defer to the observations because our parameter space is not representative of the Galactic plane, as explained above. 

%%%%%%%%%%%%%%%%%%%%%%%%%%%%%%%%%%%%%%%%%%%%%%%%%%%
%%%%%%%%%%%%%%%%%%%%%%%%%%%%%%%%%%%%%%%%%%%%%%%%%%%

\section{Applications to Galaxy Wide Observations} \label{applicationstongalaxies}
The qualitative discussion in Section \ref{numericalresults} was sufficient to understand the physics underpinning the variations of \fmol\ between our different simulated star forming regions. In this section, we make the jump from these individual star forming regions to the ISM of entire galaxies. Ideally, we would want to build a model for the ISM of a whole galaxy by appropriately summing up a number of our individual simulated clouds. To do this, we could start from observations of the molecular cloud mass function \citep[e.g.][]{2010ApJ...713..871W,2011ApJS..197...16W,2014ApJ...784....3C,2016MNRAS.tmp....9G}, however it is still highly debated whether there is a universal cloud mass function that is applicable to all galaxies, or whether the properties of clouds depend on other global physical parameters and therefore vary from galaxy to galaxy \citep[see e.g.][]{hughes13}. Given this uncertainty and as a first step, we here propose a simpler alternative method to predict how \fmol\ varies as a function of integrated galaxy properties using our simulated clouds. We make the assumption that the physical conditions found in each of our clouds, for a given set of input parameters, can represent the average physical conditions found on galaxy-wide scales for galaxies with similar physical properties. Under this assumption, a whole galaxy can be considered to be built up from an appropriate number of identical star forming regions. 

\subsection{Bayesian Inference}\label{statsmethod}
We now want (a) to parametrise an analytic prescription for how \fmol\ varies as a function of our model parameters for extragalactic observations on galaxy wide scales, and (b) to determine the minimum number of parameters needed to provide a statistically-robust fit to our data. We therefore use a Bayesian inference method to find the best fit relations and the minimum number of parameters required. Bayesian inference fitting methods have been successfully employed in several, wide-ranging, astrophysical scenarios from the derivation of the extinction law in the Perseus molecular cloud \citep{2013MNRAS.428.1606F} and Type Ia supernova light curve analysis \citep{2011ApJ...731..120M} to the extragalactic Kennicutt-Schmidt relation \citep{2013MNRAS.430..288S} and the formation and evolution of Interstellar Ice \citep{2014ApJ...794...45M}. For a more in depth explanation of the Bayesian regression fitting method we refer the reader to \citet{2007ApJ...665.1489K} and restrict ourselves here to the basic concepts. 
%From the scaling relations shown in Fig. \ref{all_scaling_relations_processed}, 
Our 3D radiative transfer methodology provides a complete model for how \cii\ varies as a function of the seven input parameters of the coupled code. However, the radiative transfer modelling is highly non-linear and complex, so we explore how well a polynomial fit can describe the outputs from the coupled 3D radiative transfer simulation, and what is the optimal number of parameters for this fit. This is done by evaluating the posterior probability of the simulated data, $y_{\mbox{\scriptsize{RT Model}}}$, given the polynomial fit, denoted $y_{\mbox{\scriptsize{QF}}}$. We assume that the measurement uncertainties associated with each of our fits, are normally distributed, therefore $y_{\mbox{\scriptsize{QF,i}}}$ is a random variable distributed like:
\begin{dmath}
\centering
y_{\mbox{\scriptsize{QF,i}}} = \mathcal{N}(y_{\mbox{\scriptsize{RT Model, i}}}, \sigma_{y_{\mbox{\scriptsize{QF, i}}}}^{2})
\label{normaldis}
\end{dmath}
where $\sigma_{y_{\mbox{\scriptsize{QF, i}}}}$ is the measurement uncertainty associated with the polynomial fit $y_{\mbox{\scriptsize{QF, i}}}$ on the $i^{th}$ model which will be an additional parameter which we need to fit. For simplicity, we assume that all $\sigma_{y_{\mbox{\scriptsize{QF,i}}}}$ are equal to the same value, $\sigma_{y_{\mbox{\scriptsize{QF}}}}$.

Under the assumption of the normal distribution in Equation \ref{normaldis}, the probability of obtaining a certain polynomial fit, given the output of the numerical modelling, combined with the fitted uncertainties and the weighting factors is:

\begin{dmath}
\centering
P(y_{\mbox{\scriptsize{QF,i}}} | y_{\mbox{\scriptsize{RT Model,i}}}, \sigma_{y_{\mbox{\scriptsize{QF}}}}) = \frac{\sqrt{\mbox{g}_{i}}}{\sqrt{2\pi \sigma_{y_{\mbox{\scriptsize{QF}}}}^2}} \times \exp \left(-\frac{\mbox{g}_{i} (y_{\mbox{\scriptsize{RT Model,i}}} - y_{\mbox{\scriptsize{QF,i}}})^2}{2\sigma_{y_{\mbox{\scriptsize{QF}}}}^2} \right) 
\end{dmath}
where $g_{i}$ is the dimensionless statistical weighting for each cloud. {\as As described in Section \ref{statweights}, weights are assigned to each of the simulated clouds to take into account how likely they are to reproduce ISM conditions typical of local galaxies. }

The next assumption to make is that all our radiative transfer simulated data points are independent, which is perfectly reasonable as we ran through each point in parameter space regardless of the other parameters. Under this assumption, all the individual probabilities can be multiplied to produce the Likelihood. By taking the log-likelihood the product returns back to a sum so:

\begin{dmath}
\centering
\label{likelihoodequation}
\mathcal{L} = -\frac{N}{2} \mbox{ln}(2\pi) -  N\mbox{ln}(\sigma_{y_{\mbox{\scriptsize{QF}}}}) - \sum\limits_{i=1}^N  \left(\frac{\mbox{g}_{i} (y_{\mbox{\scriptsize{RT Model,i}}} - y_{\mbox{\scriptsize{QF,i}}})^2}{2\sigma_{y_{\mbox{\scriptsize{QF}}}}^2} \right) + \sum\limits_{i=1}^N  \left(  \frac{\mbox{ln g}_{i}}{2}   \right)
\end{dmath}.
Maximising this log-likelihood for the polynomial fit parameters and the associated error, $\sigma_{y_{\mbox{\scriptsize{QF}}}}$, will provide us with the best fitting analytical expression alongside the one sigma error of the parametrisation.

To compare likelihoods from models with different numbers of free parameters we use two different methodologies. Firstly we employ the Akaike Information Criterion (AIC) \citep{RePEc:eee:econom:v:16:y:1981:i:1:p:3-14}: 
\begin{dmath}
\centering
AIC = -2 \mathcal{L} + 2p + \frac{2p(p+1)}{N-p-1}
\end{dmath}
where $p$ is the number of free parameters and $N$ is the sample size. The best model and the optimal number of free parameter is found by minimising the AIC. We also calculate the Bayesian Information Criterion (BIC) \citep{schwarz1978}:
\begin{dmath}
\centering
BIC = -2 \mathcal{L} + p \mbox{Log} (N)
\end{dmath}
and compare the results of both tests to ensure our results are not dependent on the choice of the information criterion used. 

\subsection{Sampling Methods and Quadratic Models}
A direct solution for the posterior probability distribution is computationally expensive and so, to efficiently and effectively sample the full parameter space, we use the well tested Python implementation of the affine-invariant ensemble sampler for Markov Chain Monte Carlo (MCMC) called emcee\footnote{An example of the code can be found at http://dan.iel.fm/emcee/current/} \citep{emcee}. 

Given the saturation effect which may occur when \cii\ is mainly emitted from the molecular regions (i.e. when \fmol\ approaches 1) and the low number of bins in our parameter space, we only fit quadratic polynomials to our data, including all second order cross-terms when multiple parameters are involved e.g for three parameters we would use:
\begin{dmath}
\centering
y_{\mbox{\scriptsize{QF,i}}} = \alpha_{1} + \alpha_{2}x_{1,i} + \alpha_{3}x^2_{1,i} + \alpha_{4}x_{2,i} + \alpha_{5}x^2_{2,i} + \alpha_{6}x_{3,i} + \alpha_{7}x^2_{3,i} + \alpha_{8}x_{1,i}x_{2,i} + \alpha_{9}x_{1,i}x_{3,i} + \alpha_{10}x_{2,i}x_{3,i}.
\end{dmath}
We also fit for $\sigma_{y_{\mbox{\scriptsize{QF}}}}$ and therefore have $2(\epsilon+1) + {^\epsilon C_2}$ free parameters to constrain, where $\epsilon$ is the number of different variables in our fits\footnote{This applies when $\epsilon$ is greater or equal to two. For one variable we have four free parameters.}. This number can range from 1 to 4 as we focus on the four input parameters of the coupled code which are also commonly-available extragalactic observables. These are the gas-phase metallicity, the electron number density of the H\small II\normalsize\space regions, the specific star formation rate and the dust mass fraction.

\subsection{Statistical Weighting Calculation}
\label{statweights}
The simulated clouds, which we now assume represent average physical conditions on galaxy wide scales, fill up a very large parameter space representing a large range of possible physical conditions. In which parts of this parameter space do galaxies actually lie? Which simulated clouds therefore represent average physical conditions in local galaxies? To account for this, we calculate a weighting factor for each cloud based on how likely it is to be representative of a local galaxy. 

To determine these weighting factors, we make use of the Herschel Reference Survey \citep[][, HRS hereafter]{2012A&A...540A..52C}, a statistically complete K-band selected sample of galaxies located between 15 and 25 Mpc \citep{2010PASP..122..261B}. We retrieve dust masses, stellar masses and star formation rates from HRS catalogs \citep{2015A&A...579A.102B,2014MNRAS.440..942C,2013A&A...550A.114B}. From these we can directly infer for 112 HRS galaxies two of the input parameters of our coupled model: dust mass fractions and specific star formation rates. Another input parameter, the electron number densities in the ionised regions, is calculated from the [SII] line intensity ratio $R=  \mbox{[SII]} \lambda6716/\lambda6731$ using the prescription of \citet{2016ApJ...816...23S}: 
\begin{dmath}
\centering
n_{e} = \sqrt{T_{e}} \left ( \frac{1.4498 - R}{0.1595R - 0.0688}    \right ) 
\end{dmath}.
where $T_{e}$ is the electron temperature and assumed to be a standard 10$^{4}$\ K, typical for H\small II\normalsize\space regions. Similar temperature assumptions have been made previously in \citet{2014MNRAS.444.3894H} and \citet{2016ApJ...816...23S}. These electrons densities are equivalent to hydrogen number densities in the H\small II\normalsize\space regions in the range of $10-1000$ cm$^{-3}$. Comparison with the input parameters for the coupled multi-phase code given in Table \ref{param_space} confirms that we have sampled the appropriate ranges to reproduce conditions typical of local galaxies. 

We bin the HRS data to match the sampling used in the radiative transfer modelling (shown in Table. \ref{param_space}), i.e for the above three parameters (dust mass fractions, SSFR and $n_e$) we bin the 112 HRS galaxies into 60 bins (5$\times$4$\times$3). It is possible to include metallicity in the binning, however, this would restrict and reduce the sample size further. If we did include metallicity our sample would now shrink to 69 objects and we would now have 120 bins, meaning that our weighting function would be comb-like leading to erroneous results as the number of bins exceed the sample size. A variant of the Freedman \& Diaconis rule \citep{freedmandiaconis} states that the number of bins must be less than the sample size, hence why we do not include metallicity into the weighting. We use the Python N-dimensional histogram routine, {\it histogramdd}, to calculate the normalised weighting for each of our simulated clouds.  

From here on, we limit our sample to clouds with cosmic ray ionisation rates equal to the average Milky Way value (10$^{-17}$ s$^{-1}$), as it is unlikely to vary much away from this value for any of the HRS galaxies, which are local normal star-forming and quiescent galaxies. Higher cosmic ray rates ($\sim$10$^{3}$x Milky Way value) are found in ULIRGs and galaxies with more enhanced star formation \citep{2015A&A...578A..70K}, hence we keep this fixed at the average Milky Way value for now. Even though the HRS sample does not have measured cosmic ray ionisation rates, based on their position in the SFR-M$_{*}$ plane, and the fact that they are local galaxies, we can be sure that they all have a value approximately equal to that of the average Milky Way cosmic ray ionisation rate. Therefore we set to 0 the statistical weight of any cloud simulated with a cosmic ray ionisation rate higher than that of the Milky Way. 

We also present in Appendix \ref{AppCR} the results if we limit our sample to clouds with cosmic ray ionisation rates equal to ten times the average Milky Way value (10$^{-16}$ s$^{-1}$). We find identical results when using the two different cosmic ray ionisation rates; using the average Milky Way value (10$^{-17}$ s$^{-1}$) does not affect our results for local universe galaxies. 
\begin{figure}
  \centering
    \includegraphics[scale=0.41]{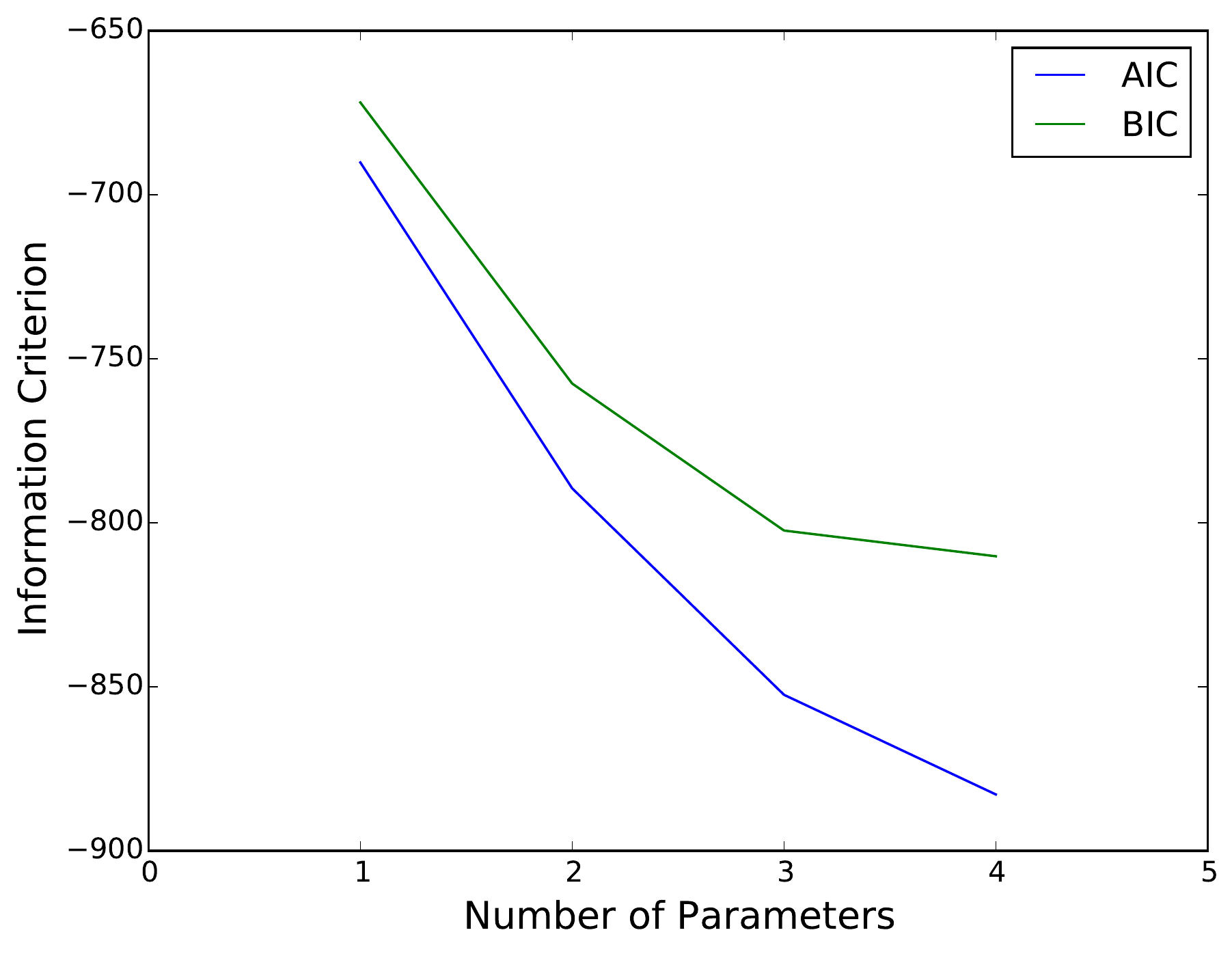}
\caption{We provide a plot for the variation of the Akaike Information Criterion and the Bayesian Information Criteria in blue and green respectively. It can be seen how, although they give different absolute numerical values (due to their different analytic expressions), they reach a minimum at four parameters.}
\label{residuals_of_final_fit} 
\end{figure}
\subsection{Statistical Results}
Using the Bayesian formalism and statistical weights described above, we fit the simulated values of \fmol\ as a function of the four key observables (density, dust mass fraction, SSFR and metallicity), allowing the number of these parameters used in any one fit to vary between 1 and 4.  As the number of free parameters increases, the quality of the fit improves as shown by both the AIC and BIC (Figure \ref{residuals_of_final_fit}).  
\begin{figure*}
  \centering
    \includegraphics[scale=0.85]{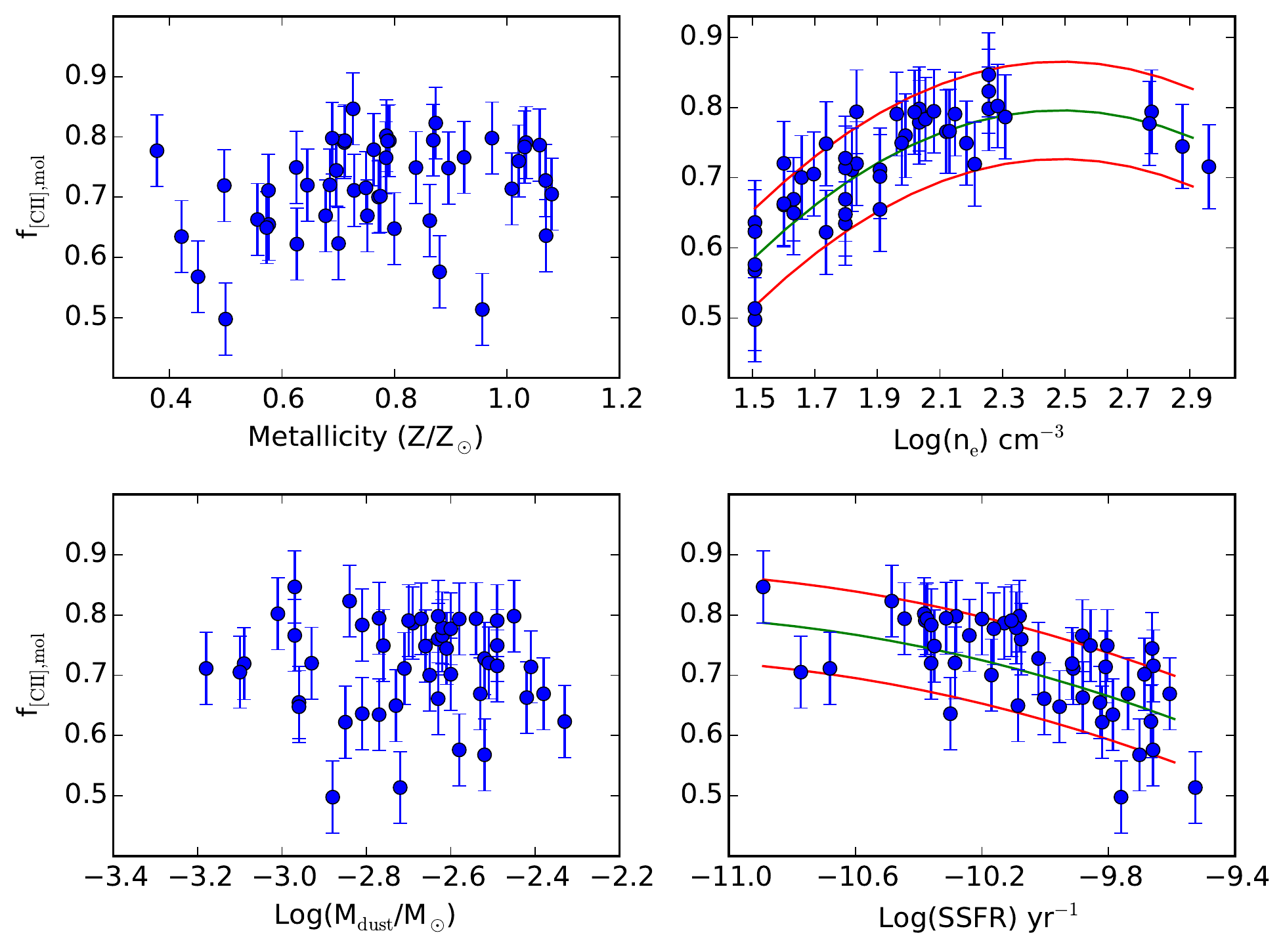}
\caption{Relation between \fmol\ as calculated from Eq. \ref{final_fit_equation_four} for galaxies from the HRS sample that have four key integrated properties (metallicity, density, dust mass fraction and SSFR). We also overlay the single prescriptions for density and SSFR, Equations \ref{final_fit_equation_one} and \ref{ssfr_quad_bayes} respectively, in the upper right and bottom right panels by the green lines. The red lines represent the one sigma errors on both prescriptions.}             
\label{hrs_sample_scaling_relations} 
\end{figure*}
We hereby present several novel prescriptions for the fraction of [C\small II\normalsize] emission emerging from molecular regions on galaxy wide scales simply involving dust mass fraction, H\footnotesize II \normalsize region electron number density, specific star formation rate and metallicity. The full analytical prescription, according to the AIC and BIC is one involving all four galaxy parameters, namely:
\begin{dmath}
\centering
f_{\rm [C\small II\normalsize],mol} = -4.405 + 0.133\frac{Z}{Z_{\odot}} - 0.172 \frac{Z}{Z_{\odot}}^{2} + 1.448\rho - 0.206\rho^{2} + 0.814\phi -0.050\phi^{2} -0.818\psi -0.032\psi^{2} -0.063\frac{Z}{Z_{\odot}}\rho +0.003\frac{Z}{Z_{\odot}}\phi -0.027\frac{Z}{Z_{\odot}}\psi -0.222\rho\phi + 0.098\rho\psi + 0.050 \phi\psi
\label{final_fit_equation_four}
\end{dmath}
where $\frac{Z}{Z_{\odot}}$ is the metallicity, $\rho = \log n_e$, $\phi =$ log $\frac{M_{dust}}{M_{*}}$ and $\psi =$ log(SSFR). The one sigma error derived from the fitting is $\sigma_{f_{\mbox{\scriptsize{[C\tiny II\scriptsize],mol}}}} =$ 0.0597 (a unit-less quantity as it's a relative fraction). Furthermore, we also present the best three, two and one parameter prescriptions as we understand acquiring all the necessary data to use Equation. \ref{final_fit_equation_four} may be a challenge. The three parameter prescription includes only dust mass fraction, $n_e$ and SSFR:
\begin{dmath}
\centering
f_{\rm [C\small II\normalsize],mol} = -3.92 + 1.50\rho - 0.209\rho^{2} + 0.471\phi - 0.072\phi^{2} -0.628\psi -0.018\psi^{2} - 0.227\rho\phi + 0.106\rho\psi + 0.027\phi\psi
\label{final_fit_equation_three}
\end{dmath}.
\begin{figure}
  \centering
    \includegraphics[scale=0.40]{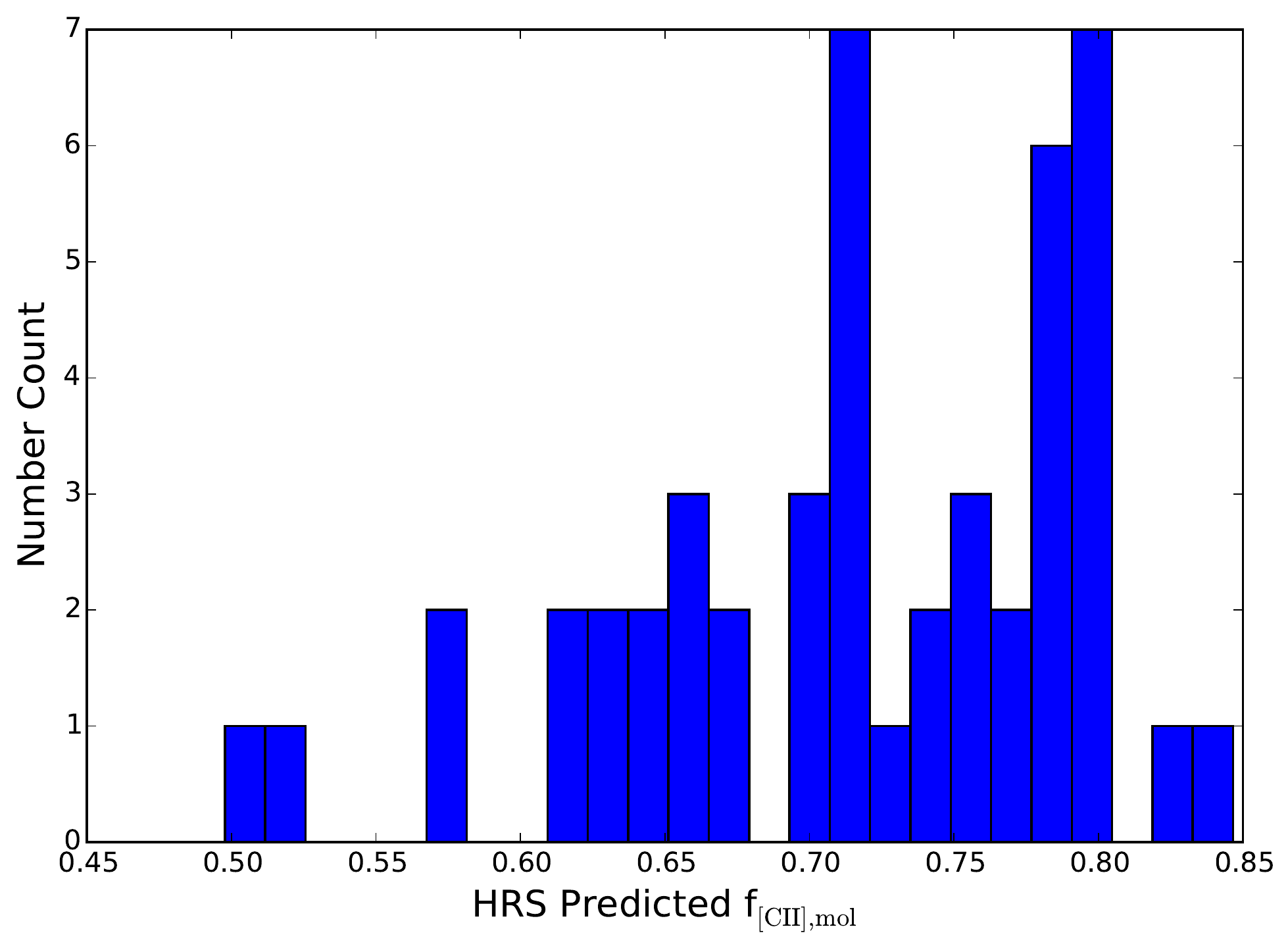}
\caption{By applying our prescription to the HRS sample we find that the majority of the galaxies have 60-80\% of their total integrated [C\scriptsize II\normalsize] emission arising from molecular regions. Due to the completeness of the sample it implies that on galaxy wide scales, in the local universe, 60-80\% of a galaxy's [C\scriptsize II\normalsize] emission will originate from molecular regions.}
\label{prescription_to_hrs} 
\end{figure}
The one sigma error in this case is is $\sigma_{f_{[C\small II\normalsize],mol}} =$ 0.061. The two parameter prescription does away with the dust mass fraction and therefore simplifies as
\begin{dmath}
\centering
f_{\rm [C\small II\normalsize],mol} = -5.63 + 1.31\rho - 0.17\rho^2 -0.87\psi - 0.034\psi^2 + 0.046\psi\rho
\label{final_fit_equation_two}
\end{dmath}
with an error of $\sigma_{f_{[C\small II\normalsize],mol}} =$ 0.064. The best fitting one parameter prescription involves only $n_e$ and has  $\sigma_{f_{[C\small II\normalsize],mol}} =$ 0.069:
\begin{dmath}
\centering
f_{\rm [C\small II\normalsize],mol} = -0.556 + 1.087\rho - 0.219\rho^2. 
\label{final_fit_equation_one}
\end{dmath}
As we will show in the next section, there is also a trend between \fmol\ and SSFR. As this latter quantity is typically more readily available to extragalactic observers than $n_e$, we also perform this one parameter fit even though it is not formally selected by the AIC and BIC.  This alternative one parameter prescription, with an associated error of $\sigma_{f_{[C\small II\normalsize],mol}} =$ 0.072, is 
\begin{dmath}
\centering
f_{\mbox{\scriptsize{[C\tiny II\scriptsize],mol}}} = -6.224 -1.235\psi - 0.0543\psi^2.
\label{ssfr_quad_bayes}
\end{dmath}
In the following section, we test and compare these five prescriptions, and then advise on the best relation to use to estimate \fmol\ for individual galaxies in Section \ref{answertopaper}.
\begin{figure}
  \centering
    \includegraphics[scale=0.45]{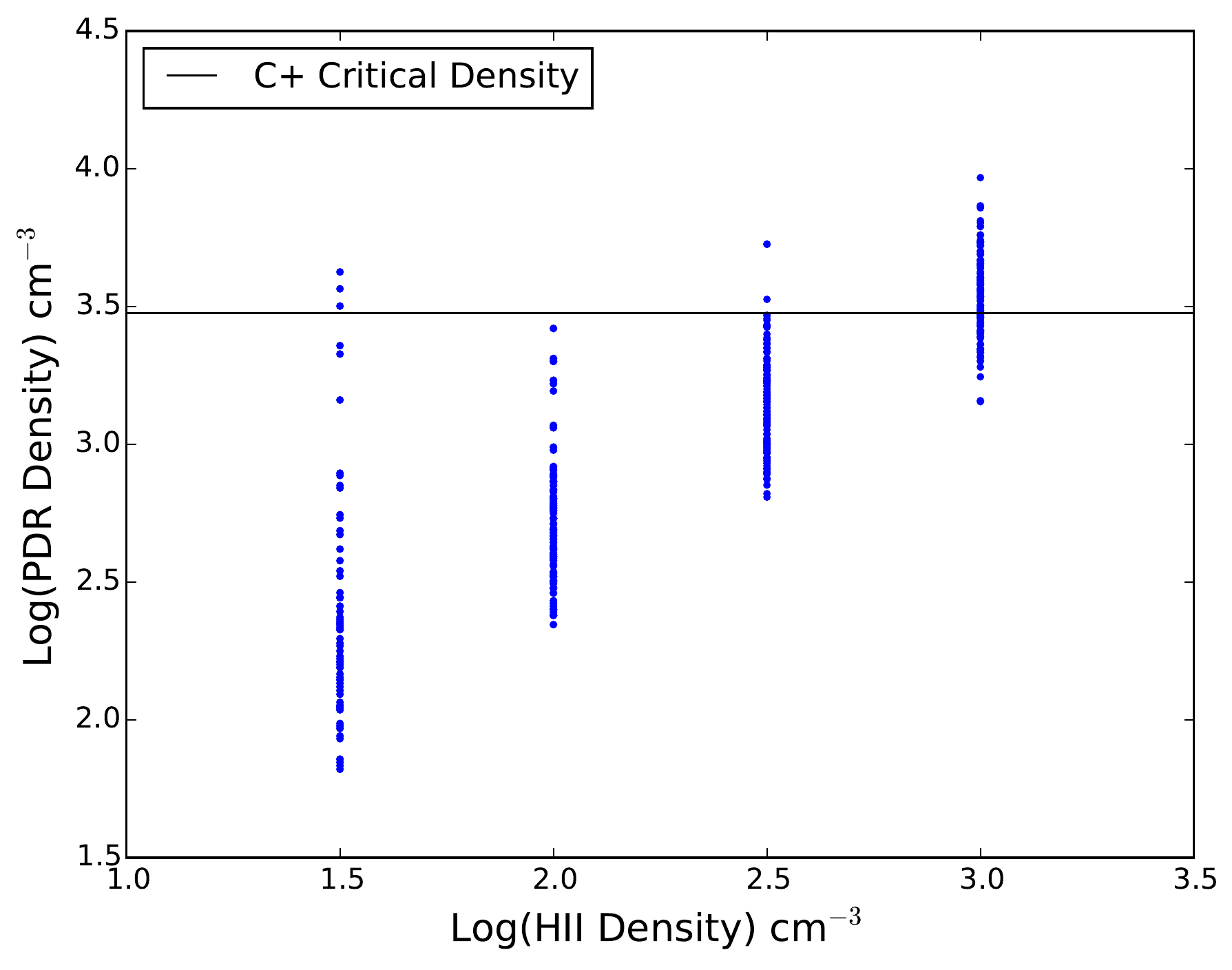}
\caption{We plot the input Hydrogen number density (equivalent to n$_{e}$) used in ionised regions against the calculated PDR number densities. As can be seen for a H\tiny II\small\space density of 10$^{3}$ cm$^{-3}$ the PDR density starts to exceed the critical density of [C\tiny II\small].}
\label{critical_density} 
\end{figure}
%%%%%%%%%%%%%%%%%%%%%%%%%%%%%%%%%%%%%%%%%%%%%%%%%%%
%%%%%%%%%%%%%%%%%%%%%%%%%%%%%%%%%%%%%%%%%%%%%%%%%%%
\subsection{Validation of the \fmol\ prescriptions and example applications}\label{obscomparison}
Measurement of the \cii\ fraction emerging from molecular regions in extragalactic objects are uncommon at best, making validating our prescription for \fmol\ against a large and complete galaxy sample impossible. However, \citet{2013A&A...554A.103P} have have measured \fmol\ across the Milky Way as part of the GOT C$^{+}$ survey.  Assuming a dust mass for the Milky Way of 10$^{7.7}$ M$_{\odot}$ \citep{2000ApJ...545L.121P}, a star formation rate of 1.65 M$_{\odot}$ yr$^{-1}$ and a total stellar mass of 10$^{10.78}$ M$_{\odot}$ \citep{2014AAS...22333604L}, a metallicity of 1Z$_{\odot}$ and finally a electron number density of 100 cm$^{-3}$, Eq. \ref{final_fit_equation_four} predicts \fmol$=75.84\pm5.97\%$ for the Milky Way.  As a comparison, the alternative prescriptions evoking fewer input parameters, Equations \ref{ssfr_quad_bayes}, \ref{final_fit_equation_one},  \ref{final_fit_equation_two} and \ref{final_fit_equation_three}, predict values of 77.6 $\pm$ 6.3$\%$, 74.42 $\pm$ 6.94$\%$, 74.42 $\pm$ 6.94$\%$ and 73.80 $\pm$ 6.10$\%$, respectively. All of these predictions are in excellent agreement with the measured value of 75$\%$ \citep{2013A&A...554A.103P}. 

Extragalactic observations have been done, however, which accurately measure the fraction of \cii\ emerging from {\it ionised} gas regions, using the [C\small II\normalsize]/[N\small II\normalsize]205$\mu$m and [N\small II\normalsize]122$\mu$m/[N\small II\normalsize]205$\mu$m ratios \citep{2006ApJ...652L.125O}.  This fraction has been measured to be between 15$\%$-65$\%$ in NGC 891 \citep{2015A&A...575A..17H} and 20$\%$-30$\%$ in the star forming region BCLMP 302 of M33 \citep{2011A&A...532A.152M}. Our ISM model here is unable to measure exactly the \cii\ fraction arising from similar ionised regions; it would have to be modified to produce the emissivity profiles across the ionised and neutral phases to discriminate the origin of \cii\ between these two phases of the ISM. We can however provide an upper limit for the fraction of  \cii\ emerging from the ionised regions as 1-\fmol$\simeq20-40\%$, in agreement with these observations. 

We also apply our prescription to the HRS galaxies which have measurements available for all four physical parameters going into Eq.  \ref{final_fit_equation_four}, and find that the typical value of \fmol\ for these representative local galaxies is 60-80\%, shown in Fig. \ref{prescription_to_hrs}. This agrees well with \citet{2016MNRAS.457.3306O} who also find \cii\ emission to be dominated by the molecular gas. Furthermore Figure \ref{hrs_sample_scaling_relations} shows how these values of \fmol\ depend on key parameters.  In the model grid (Table \ref{param_space}), all the parameters were varied independently, without enforcing any correlations between each of them. However, observations of star-forming regions and local galaxies make it clear that many of these physical properties are highly correlated. The scaling relations of Fig. \ref{hrs_sample_scaling_relations} therefore implicitly contain these physical correlations, and interestingly show no trend with metallicity and dust mass fraction. Correlations are seen however with SSFR and $n_e$ with explanations similar to those in sections \ref{sfr_explanation} and \ref{electron_density_explanation}. The initial increase in \fmol\ with $n_e$ occurs as we have already reached the critical density of \cii\ in the H\footnotesize II \normalsize region, hence emission from the H\footnotesize II \normalsize region saturates while continuing to increase in the PDR region. There is then a plateau of the relation which occurs at $\sim10^{2.5}$ cm$^{-3}$ because the densities are starting to approach the critical density of \cii\ in the PDR regions as shown in Figure \ref{critical_density}. At these densities the \cii\ emission from the PDR also saturates, leading to an overall constant value of \fmol.  
%This is to be contrasted with the scaling relations shown in Fig. \ref{all_scaling_relations_processed} which contained all the simulated clouds, regardless of their likeliness to be representative of local star-forming galaxies. 
\begin{figure}
  \centering
    \includegraphics[scale=0.45]{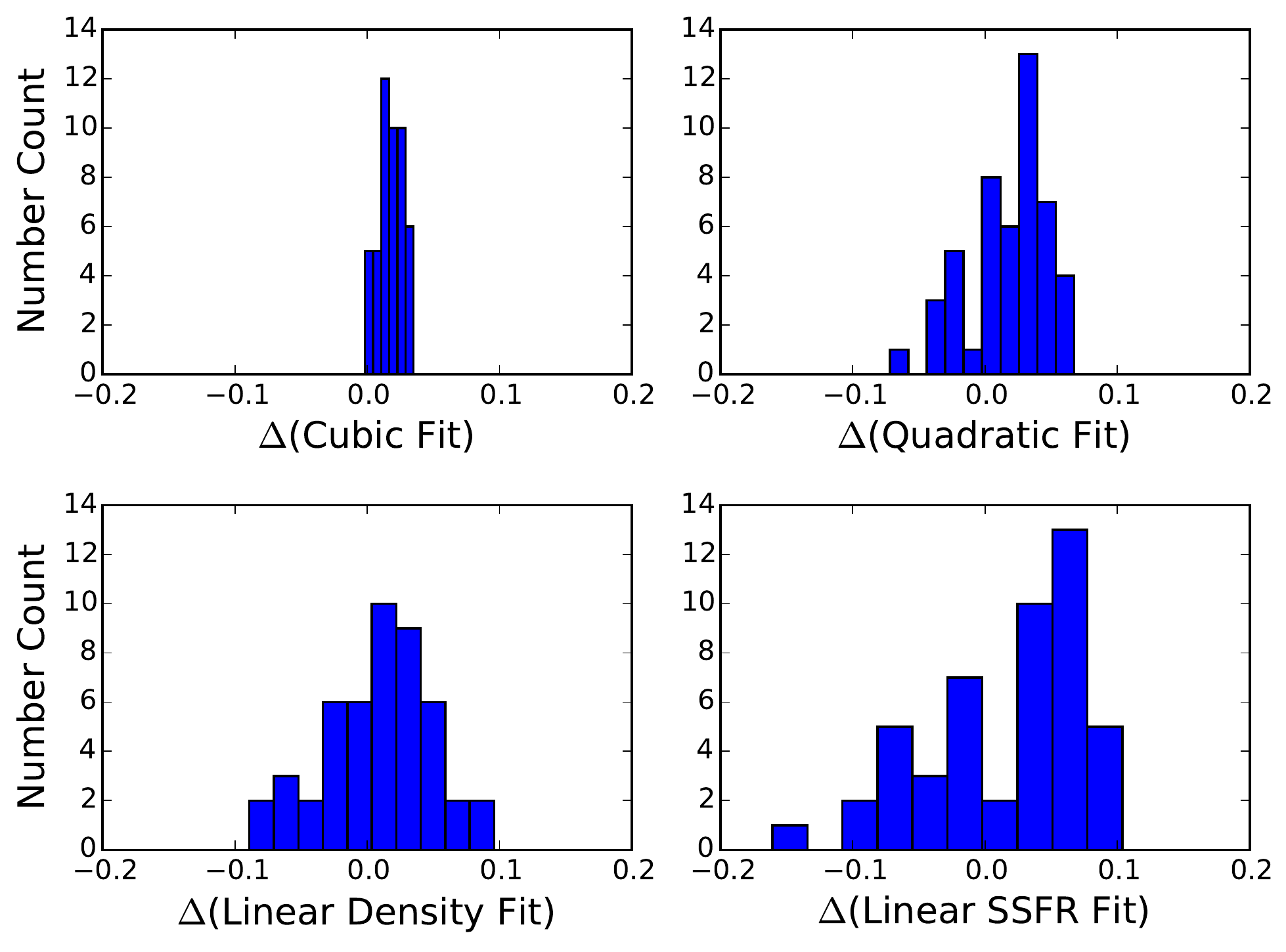}
\caption{Distribution of the offset between the best prediction of \fmol\ from Eq. \ref{final_fit_equation_four} for the HRS galaxies, and the values produced by the four alternative prescriptions (Eqs. \ref{final_fit_equation_three}-\ref{ssfr_quad_bayes}).} 
\label{hrs_comparison} 
\end{figure}
\tikzstyle{decision} = [diamond, draw, fill=blue!30, 
    text width=8.5em, text badly centered, node distance=4.25cm, inner sep=0pt, minimum height=2em]
\tikzstyle{block} = [rectangle, draw, fill=green!70, 
    text width=6em, text centered, rounded corners, minimum height=3em]
\tikzstyle{line} = [draw, -latex']
\tikzstyle{cloud} = [draw, ellipse,fill=red!70, node distance=0.0cm,
    minimum height=5em]
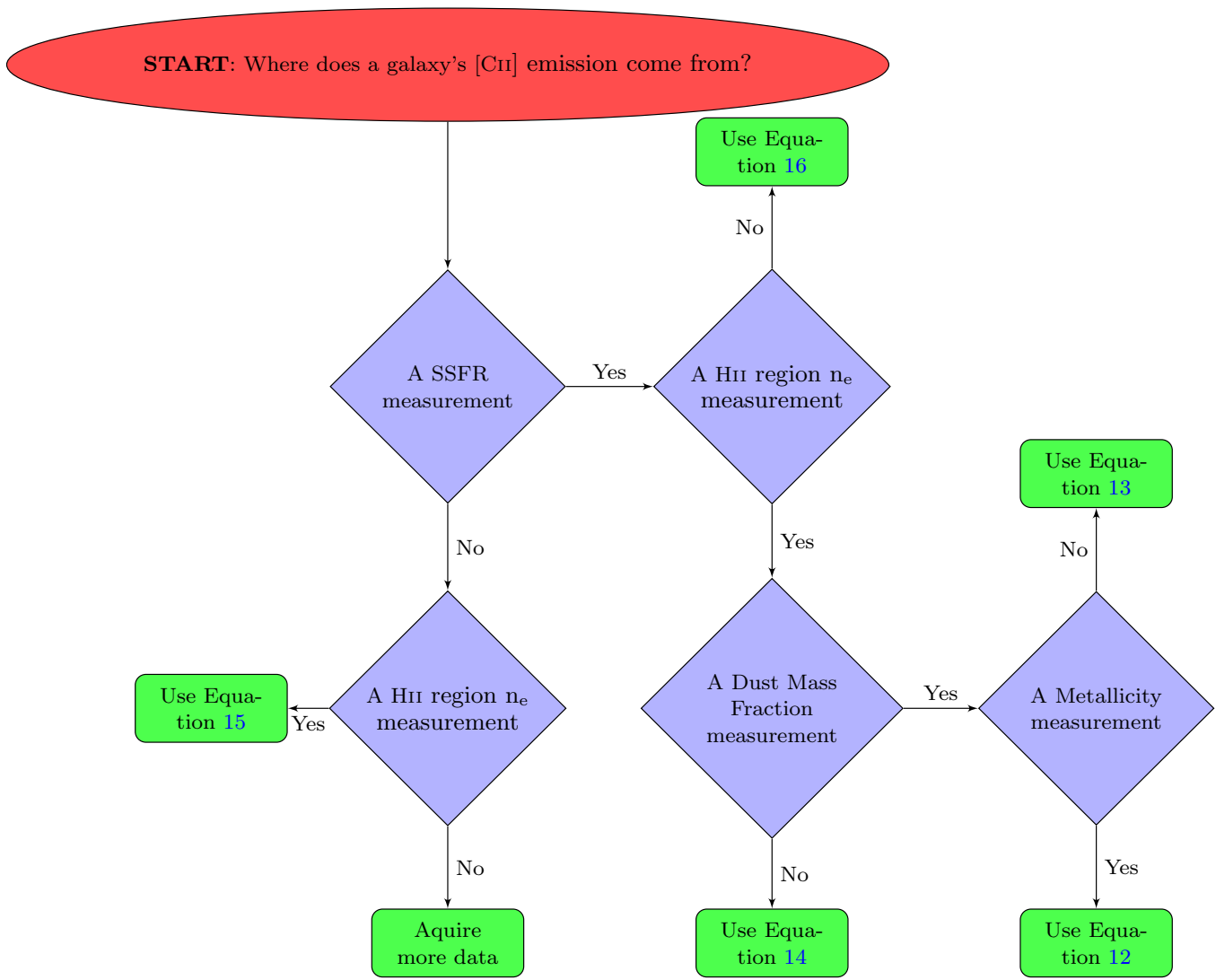
\begin {figure*}
\centering
\begin{adjustbox}{width=\textwidth}
\begin{tikzpicture}[node distance = 3.1cm, auto]
    \node [cloud] (init) {\textbf{START}: Where does a galaxy's [C\scriptsize II\normalsize] emission come from?};
    \node [decision, below of=init] (firstssfr) {A SSFR measurement};  
    \node [decision, below of=firstssfr] (seconddensity) {A H\scriptsize II\normalsize\space region n$_{\mbox{\scriptsize{e}}}$ measurement};
    \node [decision, right of=firstssfr] (thirddensity) {A H\scriptsize II\normalsize\space region n$_{\mbox{\scriptsize{e}}}$ measurement};
    \node [block, below of=seconddensity] (equationfour) {Aquire more data};
    \node [block, left of=seconddensity] (equationfive) {Use Equation \ref{final_fit_equation_one}};
    \node [block, above of=thirddensity] (equationsix) {Use Equation \ref{ssfr_quad_bayes}};
    \node [decision, below of=thirddensity] (dmfone) {A Dust Mass Fraction measurement};
    \node [decision, right of=dmfone] (metone) {A Metallicity measurement};
    \node [block, below of=dmfone] (equationseven) {Use Equation \ref{final_fit_equation_two}};
    \node [block, above of=metone] (equationeight) {Use Equation \ref{final_fit_equation_three}};
    \node [block, below of=metone] (equationnine) {Use Equation \ref{final_fit_equation_four}};
    \path [line] (init) -- (firstssfr);
       \path [line] (firstssfr) -- node {No}(seconddensity);
        \path [line] (firstssfr) -- node {Yes}(thirddensity);
         \path [line] (seconddensity) -- node {No}(equationfour);
          \path [line] (seconddensity) -- node {Yes}(equationfive);
       \path [line] (thirddensity) -- node {No}(equationsix);
       \path [line] (thirddensity) -- node {Yes}(dmfone);
       \path [line] (dmfone) -- node {Yes}(metone);
       \path [line] (dmfone) -- node {No}(equationseven);
       \path [line] (metone) -- node {No}(equationeight);
       \path [line] (metone) -- node {Yes}(equationnine);
\end{tikzpicture}
\end{adjustbox}
\caption{We here present a flowchart which can be used to constrain the fraction of [C\tiny II\small] emission from molecular regions from a galaxy. This will help to decide which equation should be used depending on which physical parameters of the galaxy have been observed and, therefore, what data is available for an individual galaxy.}
\label{flowchart} 
\end{figure*}
Finally, we test the consistency of the values of \fmol\ obtained from equations \ref{final_fit_equation_four}-\ref{ssfr_quad_bayes}.  The value of \fmol\ is calculated for the HRS galaxies using all five of these equations. Assuming that the most accurate estimate is given by the four parameter Eq. \ref{final_fit_equation_four}, the offset between the other sets of measurements and this reference are shown in Figure \ref{hrs_comparison}. As expected, the dispersion increases as the number of parameters used to calculate \fmol\ decreases, and the uncertainty on \fmol\ increases when using the one-parameter equations \ref{final_fit_equation_one} or \ref{ssfr_quad_bayes} compared to the four-parameter equation \ref{final_fit_equation_four}. This increase in uncertainty is accounted for by the larger $\sigma_{f_{[C\small II\normalsize],mol}}$ values of equations \ref{final_fit_equation_one} and \ref{ssfr_quad_bayes} versus that of equation \ref{final_fit_equation_four}.
%%%%%%%%%%%%%%%%%%%%%%%%%%%%%%%%%%%%%%%%%%%%%%%%%%%
%%%%%%%%%%%%%%%%%%%%%%%%%%%%%%%%%%%%%%%%%%%%%%%%%%%
As these relations were derived using the HRS to determine weighting factors, they are mostly applicable over the parameter space probed by the HRS galaxies which can be seen in Fig. \ref{hrs_sample_scaling_relations}. While the full parameter space covered by our simulated clouds was very large (see Tab. \ref{param_space}) some regions of this space were ignored via the weighting factors if found to be not representative of physical conditions in local galaxies. Throughout the analysis presented in Section \ref{applicationstongalaxies}, a Galactic cosmic ray ionisation rate was used, and so our prescriptions should only be used for low-redshift, normal star-forming galaxies. High-redshift galaxies, local ULIRGS, and other intensely star forming objects are very likely to have higher cosmic ray ionisation rates (100-1000$\times$ the Milky Way value). An analysis of the simulated clouds with high ionisation rates would require a representative sample of galaxies at high redshift, similar to the HRS at z $\sim$ 0, which is beyond the scope of this paper.    

Finally, throughout the modelling presented here, we held constant the N/O abundance ratio which is known to vary as a function of metallicity \citep{2011A&A...529A.149G,2008MNRAS.385.2011P}. In Appendix \ref{Appabundances} we explore how variations of this abundance ratio change our results, and find an uncertainty on \fmol\ of less than $3\%$, less than the reported errors in equations \ref{final_fit_equation_four}-\ref{ssfr_quad_bayes}); ergo this does not affect our results or the conclusions of this paper.
%%%%%%%%%%%%%%%%%%%%%%%%%%%%%%%%%%%%%%%%%%%%%%%%%%%
%%%%%%%%%%%%%%%%%%%%%%%%%%%%%%%%%%%%%%%%%%%%%%%%%%%
\section{Where does a galaxy's [C\small II\normalsize] emission come from?}\label{answertopaper}
Here we summarise the main prescriptions detailed in this paper, and provide a cookbook to help decide which prescription is appropriate for a user's specific needs given their available data. There are five equations (Equations \ref{final_fit_equation_four}, \ref{final_fit_equation_three}, \ref{final_fit_equation_two}, \ref{final_fit_equation_one}, \ref{ssfr_quad_bayes}) which accurately quantify the fraction of [C\small II\normalsize] emission emerging from molecular regions, however which one should be used? To answer this we present a flowchart in Figure \ref{flowchart}, which can be used to make this decision. The main decisions lie in determining which physical parameters of the galaxy have been observed and, therefore, what data is available. 

%%%%%%%%%%%%%%%%%%%%%%%%%%%%%%%%%%%%%%%%%%%%%%%%%%%
%%%%%%%%%%%%%%%%%%%%%%%%%%%%%%%%%%%%%%%%%%%%%%%%%%%
\subsection{Summary \& Conclusions} 
We built a new 3D multi-phase radiative transfer interface through the combination of {\sc starburst99}, {\sc mocassin} and {\sc 3d-pdr}, which can simulate all phases of the interstellar medium, from ionised to molecular, where photoionisation and photochemistry dominates. We assume pressure equilibrium between the ionised and neutral phases of the ISM, solving the thermal balance equations between the two regions to ensure self-consistency.  This interface was used to simulate a broad family of spherically-symmetric star-forming regions, with the aim of understanding how much of the total \cii\ emission originates from the cold molecular ISM under varying conditions. This is of importance for example to correctly interpret integrated \cii\ measurements for distant galaxies. 

An analysis of the relations obtained between \fmol\ and the key input parameters of the multi-phase code for these star-forming regions show that an increase in the strength of the UV radiation field (whether by increasing the current SFR or having a recent burst of star formation) leads to a decrease of \fmol. Metallicity variations can lead to both an increase and a reduction of \fmol\ depending on other global parameters, as it can both affect the cooling rate and the level of photodissociation of the CO molecule. 

To extend the analysis to the integrated \cii\ emission from extragalactic objects, we employed a Hierarchical Bayesian Inference method to identify the simulated clouds that are representative of the physical conditions in local star-forming galaxies, as found in the Herschel Reference Survey.  This is possible under the assumption that the physical conditions found in a simulated star-forming cloud can represent the average conditions found on galaxy-wide scales for objects with similar physical properties such as metallicity, SSFR, and density. 

We find that \fmol\ is best predicted using four key parameters: $n_e$, SSFR, dust mass fraction and metallicity (equation \ref{final_fit_equation_four}). We tested this prescription on the Milky Way and obtained an estimate that 75.9 $\pm$ 5.9$\%$ of its total \cii\ emission arises from molecular regions, which is in very good agreement with observations placing this number at 75$\%$ \citep{2013A&A...554A.103P}.  Given that it is relatively rare for measurements of all four of these parameters to be available for large samples of galaxies, we provide alternative prescriptions which invoke fewer parameters. These other prescriptions (equations \ref{final_fit_equation_three}-\ref{ssfr_quad_bayes}) also produce estimates for the Milky Way consistent with direct observations, although the uncertainty on \fmol\ increases slightly as the number of parameters involved in the prescription decreases. Of most practical use for many extragalactic studies is Equation \ref{ssfr_quad_bayes} which relates \fmol\ to SSFR. 

Applying the prescription to a sample of galaxies from the HRS, we find that typical galaxies in the local universe have 60-80\% of their \cii\ emission arising from molecular regions. Within this sample, \fmol\ increases with density, and decreases with SSFR. Combining the relations obtained through the bayesian analysis, we propose a decision tree in Fig. \ref{flowchart} to help determine which equation to use depending on the type of galaxy and the data products available. Using this, it is possible to estimate the relative fraction of \cii\ emerging from the molecular phase of the ISM when only galaxy-wide observations are available. 

%Correcting integrated measurements for emission from other phases of the ISM is critical to correctly interpret \cii\ emission as a star formation tracer. 

%%%%%%%%%%%%%%%%%%%%%%%%%%%%%%%%%%%%%%%%%%%%%%%%%%%
%%%%%%%%%%%%%%%%%%%%%%%%%%%%%%%%%%%%%%%%%%%%%%%%%%%
\section*{Acknowledgments}

GA would like to thank the UK Science and Technologies Facilities Council (STFC) for their support via a postgraduate Studentship, as well as Patrick J. Owen for informative discussions regarding the use of {\sc mocassin}.  AS acknowledges the support of the Royal Society through the award of a University Research Fellowship and of a Research Grant. 

\bibliographystyle{mnras}
\bibliography{modelling_paper} 

\bsp

\appendix
\begin{figure}
 \centering
    \includegraphics[scale=0.45]{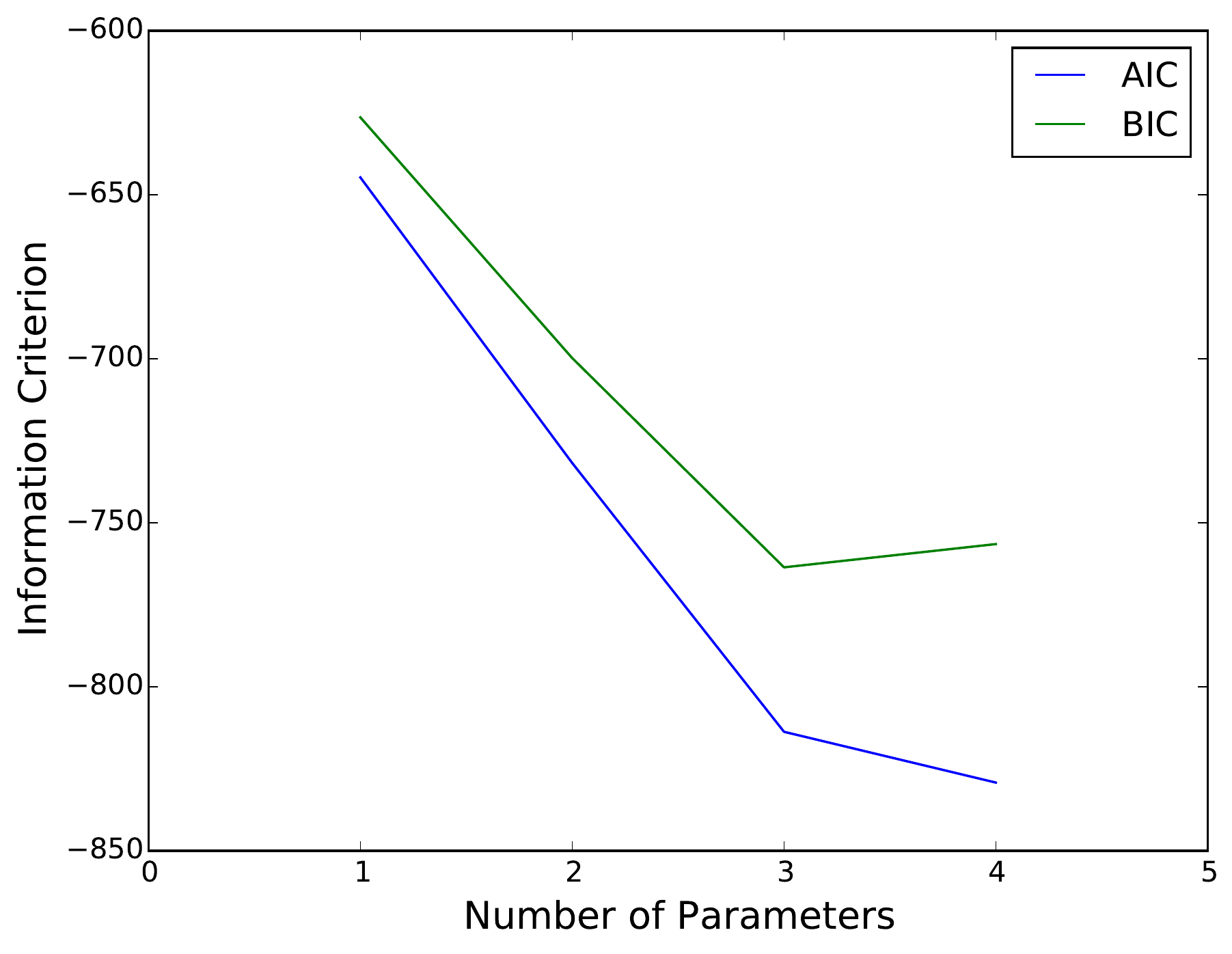}
\caption{We provide a plot for the variation of the Akaike Information Criterion and the Bayesian Information Criteria in blue and green respectively, but now for clouds with a cosmic ray ionisation rate 10x the average Milky Way value. It can be seen how they reach a minimum of three to four necessary parameters, similar to Fig. \ref{residuals_of_final_fit} .}
\label{aic_bic_relations_e16} 
\end{figure}

%%%%%%%%%%%%%%%%%%%%%%%%%%%%%%%%%%%%%%%%%%%%%%%%%%%
%%%%%%%%%%%%%%%%%%%%%%%%%%%%%%%%%%%%%%%%%%%%%%%%%%%
%%%%%%%%%%%%%%%%%%%%%%%%%%%%%%%%%%%%%%%%%%%%%%%%%%%
%%%%%%%%%%%%%%%%%%%%%%%%%%%%%%%%%%%%%%%%%%%%%%%%%%%
\section{Varying the cosmic ray ionisation rate} \label{AppCR}
\begin{figure}
 \centering
    \includegraphics[scale=0.40]{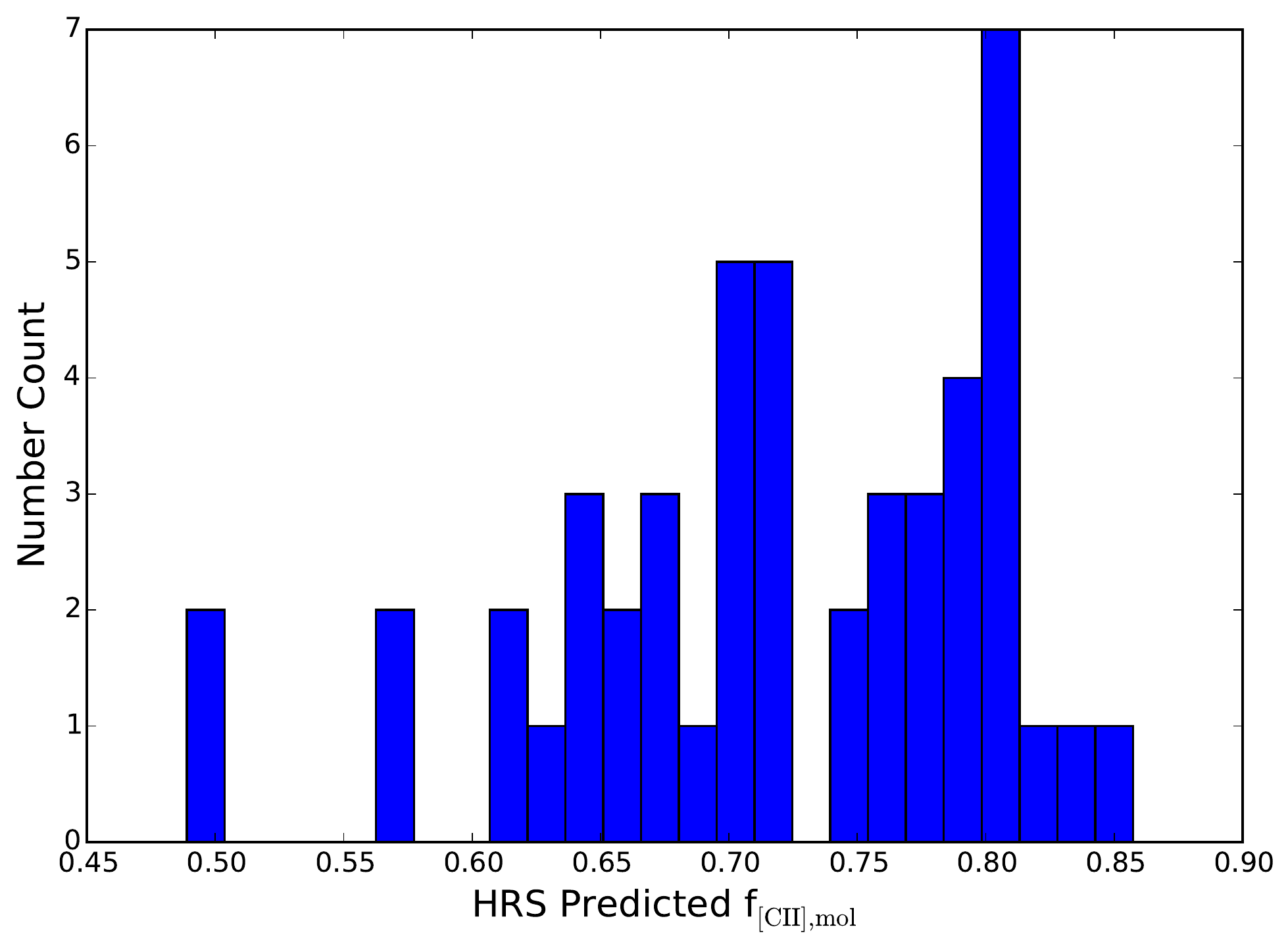}
\caption{From the HRS sample we find that the majority of the galaxies have 60-80\% of their total integrated [C\tiny II\small] emission arising from molecular regions, even for a cosmic ray ionisation rate 10x the average Milky Way value, a similar result to that in Fig. \ref{prescription_to_hrs}.}
\label{prescription_to_hrs_e16} 
\end{figure}

We performed a similar analysis as in Section \ref{statsmethod} for clouds with a cosmic ray ionisation rate 10x the average Milky Way value (10$^{-16}$ s$^{-1}$). We find that the same four parameters emerge as necessary to provide the a prescription for \fmol. We use this higher cosmic ray ionisation rate and produce identical plots to Figs \ref{residuals_of_final_fit}, \ref{hrs_sample_scaling_relations} and \ref{prescription_to_hrs}.
\begin{figure*}
 \centering
    \includegraphics[scale=0.85]{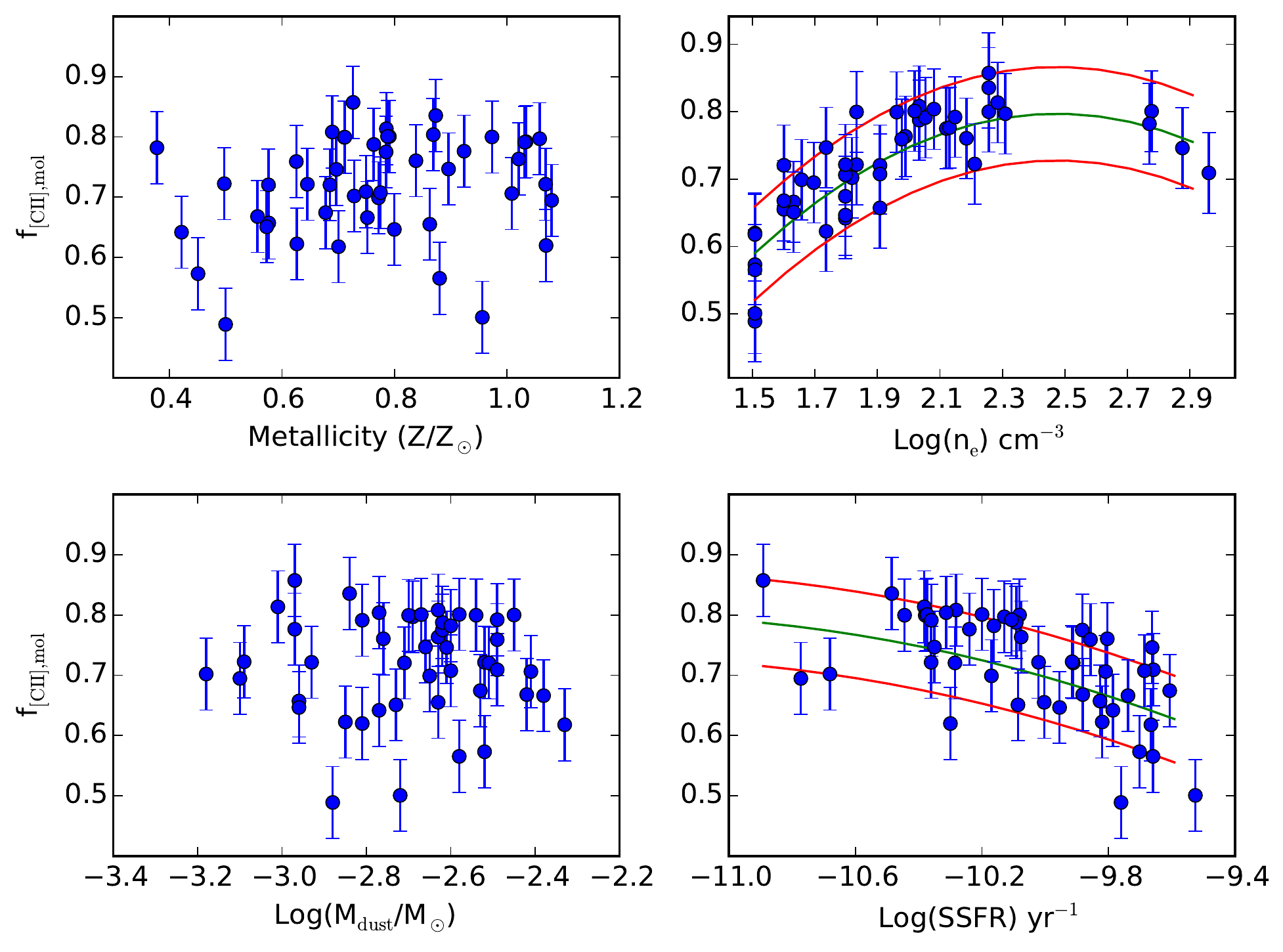}
\caption{Relation between \fmol\ for galaxies from the HRS sample for a cosmic ray ionisation rate 10x the average Milky Way value, a comparable plot to Fig. \ref{hrs_sample_scaling_relations}. The different coloured lines here represent the same as those in Fig. \ref{hrs_sample_scaling_relations}.}
\label{hrs_sample_scaling_relations_e16} 
\end{figure*}
We show in Fig \ref{aic_bic_relations_e16} a plot for the AIC and BIC, similar to Fig. \ref{residuals_of_final_fit}, and find three to four parameters are needed, similar to that in Section \ref{statsmethod}. Once we obtained our prescription we applied it to the HRS objects, which have observed values for the four important parameters. We present these results in Figure. \ref{hrs_sample_scaling_relations_e16} and, qualitatively, it is clear that the results are very similar and almost identical to those shown in Figure. \ref{hrs_sample_scaling_relations}. Finally we bin the HRS results, similar to Fig. \ref{prescription_to_hrs}, and again find that, even for clouds with a cosmic ray ionisation rate 10x the average Milky Way value, the majority of the galaxies have 60-80\% of their total integrated [C\scriptsize II\small] emission arising from molecular regions, shown in Fig. \ref{prescription_to_hrs_e16}.

Overall we claim that, even if a galaxy is thought to have cosmic ray ionisation rates ten times larger than the Milky Way value, our prescriptions detailed in Section \ref{applicationstongalaxies} are still robust and accurate as the cosmic ray ionisation rate value does not affect the results at these levels. Higher redshift objects and ULIRGS will have cosmic ray ionisation rates more than $\sim$10$^{3}$ times that of the Milky Way, and for those cases our prescription, in Section \ref{applicationstongalaxies}, starts to break down. Furthermore our prescription would not be valid at high redshift because the complete HRS sample is only complete for the low redshift universe. Therefore, our prescription is accurate only for low redshift, star forming and quiescent, galaxies regardless of their cosmic ray ionisation rate.

Obtaining a similar prescription for high redshift objects is possible, using the above method, however a statistically complete sample of galaxies at high redshift would be needed to provide the weightings necessary for the Bayesian Inference method. This could be done via a Machine Learning technique to generate a predictive sample of galaxies at high redshift, however this is beyond the scope of this paper.

\begin{table*}
\centering
  \caption{To test for the effect of varying the N/O ratio as a function of metallicity we re-run three of our {\sc mocassin} runs. We find that, even in the most extreme cases, our calculations could underestimate the [C\scriptsize II\small] emission from the ionised regions by 3.7-7.7\%}
  \begin{tabular}{@{}llllll@{}}
  \hline
   Metallicity     & Log(SSFR) & n$_{e}$  &  [CII] with &  [CII] with & Fractional   \\   
   &   & &  N/O constant &  N/O varying & difference\\
   (Z/Z$_{\odot}$)  & (yr$^{-1}$) &(cm$^{-3}$) &  (L/L$_{\odot}$) &  (L/L$_{\odot}$) & (\%)\\
 \hline
 0.2 & -11.5 & 10$^{1.5}$ & 5.15$\times$10$^{-3}$ & 5.19$\times$10$^{-3}$  & 7.7 \\
 0.2 & -10.5 &10$^{2.0}$ & 11.32$\times$10$^{-3}$ & 11.92$\times$10$^{-3}$ & 5.3\\
 0.2 & -10.5 & 10$^{3.0}$ & 9.55$\times$10$^{-3}$ & 9.91$\times$10$^{-3}$ & 3.77\\
 \hline
\end{tabular}
\label{NOtest}
\end{table*}

\section{Varying chemical abundances}\label{Appabundances}
When varying the metallicity parameter, we scaled all the abundances in Table \ref{abundances} equally, except for hydrogen and helium. This means that the relative abundances between non-hydrogen and helium elements is constant. While this is generally correct, it is not true of the N/O ratio, which varies as a function of metallicity \citep{2011A&A...529A.149G,2008MNRAS.385.2011P}.  When $\log ({\rm O/H})+12>8.2$, nitrogen is a secondary element and the N/O ratio decreases with metallicity.  However, when $\log ({\rm O/H})+12<8.2$, nitrogen is a primary element and the N/O ratio is of constant value 10$^{-1.5}$. Therefore, by assuming a constant N/O ratio, we have over-supplied the low metallicity clouds with nitrogen, which could lead to an erroneous [C\tiny II\small] emission calculation in the {\sc mocassin} simulations. Given that nitrogen is a coolant in ionised regions only, this would manifest itself as an underestimation of the [C\tiny II\small] emission from the ionised regions. 

To test for the effect of the varying N/O ratio with metallicity, we adopt the prescription from \citet{2008MNRAS.385.2011P} and re-run {\sc mocassin} for three of the low metallicity clouds ($Z=0.2 Z_{\odot}$) with different SSFR and $n_e$ (the age of the secondary burst is kept constant). We find that assuming a constant N/O could lead to an underestimation of the [C\tiny II\small] emission from the ionised of only 3.7- 7.7\%. Since ionised regions contribute between 20-40\% of the total [C\tiny II\small] emission (Sec. \ref{obscomparison}), this corresponds to an uncertainty on \fmol\ of less than $3\%$, less than the reported errors in equations \ref{final_fit_equation_four}-\ref{ssfr_quad_bayes}. This is therefore not a dominant source of uncertainty; the results for these runs are shown in Table \ref{NOtest}.

\end{document}